\documentclass[times,astrobib,amssymb]{mn2e}
\usepackage{psfig}
\usepackage{graphicx,color}
\newcommand{\lapp}{\mbox{\raisebox{-0.3em}{$\stackrel{\textstyle <}{\sim}$}}}
\newcommand{\gapp}{\mbox{\raisebox{-0.3em}{$\stackrel{\textstyle >}{\sim}$}}}
\newcommand{\be}{\begin{equation}}
\newcommand{\en}{\end{equation}}

\def\zem{$z_{\rm em}$~}

\def\mgii{Mg~{\sc ii} }

\def\mgiia{Mg~{\sc ii}$\lambda$2796}

\def\mgiiab{Mg~{\sc ii}$\lambda\lambda$2796,2803}

\def\caii{Ca~{\sc ii} }
\def\caiia{Ca~{\sc ii}$\lambda$3935}
\def\caiib{Ca~{\sc ii}$\lambda$3970}
\def\nai{Na~{\sc i} }
\def\naia{Na~{\sc i}$\lambda$5889}
\def\naib{Na~{\sc i}$\lambda$5895}

\def\kms{km~s$^{-1}$}
\title[QSO-galaxy pairs]
{GMRT mini-survey to search for 21-cm absorption in Quasar-Galaxy Pairs at $z\sim0.1$}
\author[Gupta et al.]{N. Gupta$^{1}$\thanks{E-mail: Neeraj.Gupta@csiro.au}, R. Srianand$^{2}$, 
D.V. Bowen$^{3}$, D.G. York$^{4}$ and Y. Wadadekar$^{5}$ \\
$^{1}$ Australia Telescope National Facility, CSIRO, Epping, NSW 1710, Australia \\
$^{2}$ Inter University Centre for Astronomy and Astrophysics, Post Bag 4, Ganeshkhind, Pune 411 007, India \\
$^{3}$ Department of Astrophysical Sciences, Peyton Hall, Princeton University, Princeton NJ 08544, USA \\
$^{4}$ Department of Astronomy \& Astrophysics and Enrico Fermi Institute, 5640 S. Ellis Avenue, 
Univ. of Chicago, Chicago, IL 60637, USA \\
$^{5}$ National Centre for Radio Astrophysics, Post Bag 3, Ganeshkhind, Pune 411 007, India \\
}

\begin{document}

\date{Accepted. Received; in original form }

\pagerange{\pageref{firstpage}--\pageref{lastpage}} \pubyear{2010}

\maketitle

\label{firstpage}

\begin{abstract}
We present the results from our 21-cm absorption survey of a sample 
of 5 quasar-galaxy pairs (QGPs), with the redshift of the galaxies in
the range 0.03$\le z_g \le 0.18$, selected from the Sloan Digital Sky Survey 
(SDSS). The H~{\sc i} 21-cm absorption was searched towards the 9 sight lines 
with impact parameters ranging from $\sim$10 to $\sim$55\,kpc using 
Giant Metrewave Radio Telescope (GMRT).  
21-cm absorption was detected only in one case i.e. towards the Quasar 
($z_q$ = 2.625 SDSS\,J124157.54+633241.6) $-$ galaxy 
($z_g$ = 0.143 SDSS\,J124157.26+633237.6) pair with the impact parameter 
$\sim$11\,kpc.  The quasar sight line in this case pierces through the stellar 
disk of  a galaxy having near solar metallicity (i.e (O/H)+12 = 8.7)
and star formation rate uncorrected for dust attenuation of 0.1\,M$_\odot$ yr$^{-1}$.
The quasar spectrum reddened by the foreground galaxy is well fitted with 
the Milky Way extinction curve (with an A$_V$ of 0.44) and the estimated 
H~{\sc i} column density is similar to the value obtained from 21-cm 
absorption assuming spin temperature (T$_{\rm S}$) of 100\,K. 

In the remaining cases, our GMRT spectra provide upper limit on $N$(H~{\sc i}) in the range,
(10$^{17} - 10^{18}$)$\times$T$_{\rm S}$\,cm$^{-2}$. 
Combining our sample
with the $z\le0.1$ data available in the literature, we find the detectability
of 21-cm absorption with integrated optical depth greater than 0.1\,km\,s$^{-1}$ 
to be 50\% for the impact parameter less than 20\,kpc. 
Using the surface brightness profiles and
well established relationship between the optical size and extent of
the H~{\sc i} disk known for {\sl nearby} galaxies, we conclude that in most
of the cases of 21-cm absorption non-detection,
the sight lines may not be passing through the H~{\sc i} gas 
(1$\sigma$ column density of few times 10$^{19}$\,cm$^{-2}$).  
We also find that in comparison to the absorption systems 
associated with these QGPs, 
$z<1$ DLAs with 21-cm absorption detections
have lower \caii equivalent widths despite having higher
21-cm optical depths and smaller impact parameters.  
This suggests that the current sample of 
DLAs may be a biased population that 
avoids sight lines through dusty star-forming galaxies.  
A systematic survey of QGPs over a wider redshift 
range using a large sample is needed to confirm these findings and 
understand the nature of 21-cm absorbers. 
\end{abstract}
%

\section{Introduction}
Observations of galaxies via 21-cm line, both emission and 
absorption, have proven to be valuable for our 
understanding of the abundance and distribution of  
H~{\sc i} gas, and processes leading to star formation.  
In the local Universe, blind 21-cm emission-line surveys 
using single-dish telescopes provide reliable measurements 
of the H~{\sc i} mass density (Zwaan et al. 2005).  
These surveys are complemented with the spatially 
resolved H~{\sc i} images obtained using interferometers 
to trace the large scale dynamics of 
the galaxies
(e.g. Bosma 1981;  
Fisher \& Tully 1981; Meyer et al. 2004; 
Oosterloo, Fraternali \& Sancisi 2007; 
de Blok et al. 2008; Walter et al. 2008). 
Since H~{\sc i} gas in galaxies extends farther 
than the stellar disk, it is also the component of 
the galaxy that is most affected by interactions 
and is therefore a good tracer of tidal interactions 
and merger effects 
(Haynes, Giovanelli \& Robert 1979; 
Rosenberg \& Schneider 2002).
However, beyond the local Universe (i.e. $z\gapp0.2$) 
21-cm emission is not easily 
detectable with current radio telescopes 
(Verheijen et al. 2007; Catinella et al. 2008). 
Therefore, H~{\sc i} 21-cm emission line observations 
need to be complemented with absorption 
line studies to trace the evolution of the atomic gas 
component of galaxies.  

Unlike 21-cm emission, detectability of 21-cm absorption 
is not limited by distance and depends only on the 
strength of the background radio sources and 21-cm absorption 
cross-section of the galaxies.  
Due to the limitations imposed by the narrow receiver bandwidths and 
the hostile radio-frequency interference (RFI) environment, 
systematic blind searches of 21-cm absorption have not 
been possible till now.   
Consequently, a host of absorption-line surveys have 
been designed to detect 21-cm absorption from the 
gas pre-selected by the presence of a 
damped Lyman-$\alpha$ absorber (DLA) or Mg~{\sc ii} 
absorption with a view to trace physical conditions in the 
interstellar medium of galaxies at $0<z\le3.5$
(e.g. Briggs \& Wolfe, 1983; Carilli et al. 1996; Lane 2000; 
Kanekar \& Chengalur 2003, 2009a; Curran et al. 2007; 
Gupta et al. 2007, 2009; Srianand et al. 2010).
%
%
\begin{table*}
\caption{Observing log for the GMRT observations.}
\begin{center}
\begin{tabular}{ccccccccc}
\hline
\hline
Quasar              &  $z_q$  & $r$ mag& Galaxy             &  $z_g$ & Date       &Channel & Time       \\
                    &         &        &                    &        &            &width   &            \\
                    &         &        &                    &        &            &(km\,s$^{-1}$)&(hrs) \\
(1)                 &   (2)   &  (3)   &  (4)               &  (5)   & (6)        &(7)     &(8)         \\
\hline                                                                                                
J082153.82+503120.4 & 2.130   & 17.8   &J082153.75+503125.7 & 0.1835 & 2007 Sep 19&3.9     &8.0         \\
J084957.97+510829.0 & 0.584   & 18.2   &J084957.48+510842.3 & 0.0734 & 2007 Jun 19&3.5     &7.7         \\
J111023.85+032136.1 & 0.966   & 18.6   &J111025.09+032138.8 & 0.0301 & 2008 Mar 01&3.4     &7.6         \\
J122847.42+370612.0 & 1.517   & 18.2   &J122847.72+370606.9 & 0.1383 & 2008 Mar 02&3.7     &7.5         \\
J124157.54+633241.6 & 2.625   & 17.9   &J124157.26+633237.6 & 0.1430 & 2008 Mar 03&3.8     &7.7         \\
                    &         &        &                    &        & 2009 Jun 13&3.8     &8.0         \\
                    &         &        &                    &        & 2009 Jun 14&3.8     &5.4         \\
\hline
\end{tabular}
\end{center}
\begin{flushleft}
Column 1: Quasar name. 
Column 2 and 3: Quasar redshift and $r$-band magnitude respectively. 
Column 4: Galaxy name. Column 5: Galaxy redshift. 
Column 6: Date of observation. Column 7: channel width in km\,s$^{-1}$.
Column 8: time on source in hrs. \\
\end{flushleft}
\label{gmrtlog}
\end{table*}

Although remarkable progress has been 
achieved in these emission and absorption line studies, 
an important missing link is the lack of understanding of 
the nature of absorption line galaxies required to establish 
a 
direct connection with the nearby galaxies detected in 
H~{\sc i} 21-cm emission. 
At $z\lapp2$, the most successful search for 21-cm absorbers, to date, 
has resulted from the Giant Metrewave Radio Telescope (GMRT) survey 
based on the sample of {\sl strong} \mgii absorbers 
(rest equivalent width, W$_{\rm r}$(\mgiia)$\ge$1.0\AA) 
selected from the Sloan Digital Sky Survey (SDSS) database  
(Gupta et al. 2009).  
Connections between \mgii
absorbers and galaxies at these redshifts are established
by either searching for the galaxy responsible for the \mgii absorption in 
QSO spectra 
or by searching for \mgii absorption lines at the redshifts of known galaxies 
in the spectra of background QSOs.
Based on the first approach, it is found that the success rate 
of detecting L$_{\rm *}$ galaxies at the same redshift as the known
strong \mgii absorbers is close to unity (e.g. Bergeron \& Boiss\'e 1991; 
Steidel 1995). 
The second approach shows that galaxies
close to QSO sight lines do not always produce detectable absorption
(see Bechtold \& Ellingson 1992; Tripp \& Bowen 2005). The existing
observations are consistent with \mgii absorption originating
from extended gas halos with covering factors much less than 
unity (see Churchill et al. 2005; Kacprzak et al. 2008).
The presence of large number of high quality QSO spectra 
in the SDSS database allows 
(i) the direct detection of emission lines from these galaxies 
superimposed on the QSO spectra (Quashnock et al. 2008; Noterdaeme, Srianand
\& Mohan 2010; Borthakur et al. 2010) and (ii) detection of emission lines
in the stacked spectra (Noterdaeme et al. 2010;
M\'enard et al. 2009). These observations are consistent with
strong \mgii systems being produced by gas
within $\sim$10\,kpc of a star-forming galaxy. 
%

%
The 21-cm optical depth, in addition to H~{\sc i} 
column density, $N$(H~{\sc i}), also depends on the 
spin temperature (T$_{\rm S}$). As, to begin with, 
there is no direct relationship between 
W$_{\rm r}$(Mg~{\sc ii}) and $N$(H~{\sc i})  
it is difficult to use the results of Mg~{\sc ii} systems 
alone to establish the connection between 21-cm 
absorbers and galaxies.
A systematic survey of 21-cm absorption in a sample 
of bright radio sources with sight lines passing through the 
gas disks/halos of foreground galaxies is required to 
determine the 21-cm absorption cross-section of the galaxies. 
The availability of large number of spectra in the SDSS database 
allows us to build a complete sample of associations of 
radio sources and galaxies suitable for this purpose. 
We refer to these fortuitous associations as quasar-galaxy pair (QGP) 
if the background radio source is a quasar, and 
radio source-galaxy pair (RGP) if the background source is 
a radio galaxy. 
As a prelude to our systematic survey to search 
for 21-cm absorption in a complete sample of QGPs and RGPs 
(foreground galaxy redshift $z_g\lapp$0.3) 
selected from SDSS database, we have searched 
for 21-cm absorption in 5 QGPs as a pilot project with GMRT.    
The purpose of this paper is to present the results 
from the observations of these 5 QGPs. 
In Section~2, we present the details of GMRT observations and 
data analysis.
The details of QGPs, their radio and optical properties, and 
results from the GMRT observations are presented in 
Section~3. 
In Section~4, we discuss the detectability of 21-cm absorption 
by combining our sample with the data from the literature. 
Results are summarised in  Section~5.  
Throughout this paper we use the $\Lambda$CDM cosmology with 
H$_o$=71\,km\,s$^{-1}$Mpc$^{-1}$, $\Omega_m$=0.27 and $\Omega_\Lambda$=0.73 (Spergel et al. 2007).

%
\section{Observations and data reduction}
Our sample of 5 QGPs presented in Table\,\ref{gmrtlog} is drawn from the 
SDSS DR5 database. 
The sample was constructed by cross-correlating SDSS DR5 
photometric and spectroscopic database 
for QSOs and galaxies  
with the FIRST and NVSS surveys to identify  
QGPs with background quasar flux density at 1.4\,GHz in 
excess of $\sim$50\,mJy.   
Redshifts for the galaxies for 4 QGPs presented here were 
obtained from the publicly available SDSS spectra.
For the remaining quasar ($z_q$ = 2.130 SDSS\,J082153.82+503120.4) $-$
galaxy ($z_g$ = 0.184 SDSS J082153.75+503125.7) pair, 
we have estimated the galaxy redshift using our own 
Apache Point Observatory (APO) 3.5-meter telescope observations. 
We also obtained the APO 3.5-meter telescope observations for 
quasar($z_q$ = 0.584 SDSS\,J084957.97+510829.0) $-$
galaxy ($z_g$ = 0.073 SDSS\,J084957.48+510842.3) pair 
to complement the SDSS spectrum.    
Optical parameters used in this study are 
extracted from the SDSS photometric and spectroscopic  
database and the data obtained from our 
APO observations.  Details of these optical data are provided 
in the subsequent sections. 

We observed our sample of 5 QGPs using the GMRT L-band receiver.
For these observations we used the 2\,MHz baseband bandwidth split into 128 frequency
channels (spectral resolution $\sim$4\,km\,s$^{-1} $ per channel) centered at the 
redshifted H~{\sc i} frequency estimated from galaxy redshifts. 
The observing log is presented in Table~\ref{gmrtlog}.
Data were acquired in the two linear polarization channels called XX and YY.
We observed the standard flux density calibrators 3C\,48, 3C\,147 and 3C\,286 for 10-15\,mins
every 2-3\,hrs to obtain reliable flux and bandpass calibration.  A phase calibrator was
also observed for 10\,mins every $\sim$45\,mins.
GMRT data were reduced using the NRAO Astronomical Image Processing System
(AIPS) following the standard procedures.
After the initial flagging and calibration, source and calibrator data were
examined to flag and exclude the baselines and timestamps affected by RFI. 
Applying complex gains and bandpass obtained
using the flux and phase calibrators, a continuum map of the source was made
using absorption-free channels. Using this map as a model, self-calibration
complex gains were determined and applied to all the frequency channels.  The
same continuum map was then used to subtract the continuum emission from
the dataset. This continuum subtracted dataset was then imaged separately
for the channels XX and YY to get 3-dimensional (third axis being the 
frequency) data
cubes.
The spectra at the location of radio components of interest were extracted
from these cubes and compared for consistency.
If necessary a first-order
cubic-spline was fitted to remove the residual continuum from the spectra.
The two polarization channels were then combined to get the final stokes I
spectrum, which was then shifted to the heliocentric frame.
%
\begin{figure}
\centerline{{
\includegraphics[scale=0.5]{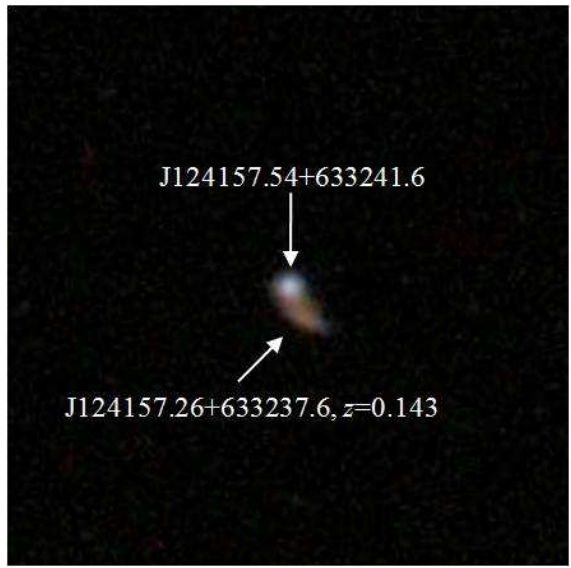}
}}
\caption[]{Color representation of the 
quasar ($z_q = 2.625$ SDSS J124157.54+633241.6)$-$galaxy 
($z_g$= 0.143  SDSS J124157.26+633237.6) pair.
The image is 77$^{\prime\prime}$ on a side, with north-east 
towards the top left corner, and centered on the QSO. 
}
\label{j1241sdsscol}
\end{figure}

\section{Results}
This survey of 5 QGPs resulted in the detection of 21-cm absorption from  
one QGP whereas no 21-cm absorption was 
detected for the remaining four QGPs.  Results are summarised 
in the Table~\ref{gmrtres}. In the following,  
we present results for the individual QGPs.

\begin{table*}
\caption{Results of GMRT observations and other parameters derived from the SDSS data.}
\begin{center}
\begin{tabular}{lcccccccc}
\hline
\hline
Quasar       & Peak       & Spectral& $\tau$ & $\int\tau$dv&Petrosian                    & Angular   & Impact     & W$_r$(\caiia) \\
             & flux       &  rms    &        &             &Radius (R$_{90}$)   & Separation& parameter  &          \\
             & (mJy)      &(mJy/b/ch)& &(km\,s$^{-1}$)     &($^{\prime\prime}$) & ($^{\prime\prime}$)  &  (kpc)   & (\AA)  \\
 (1)         &  (2)   &  (3) & (4)    & (5)   &    (6)     &   (7) & (8) &(9) \\
\hline
J0821+5031   & 47.8   &   0.9 &$<$0.019 &$<$0.38  & 3.48 & 5.2     & 15.9  & $<$0.25    \\ 
J0849+5108   & 248    &   1.0 &$<$0.004 &$<$0.08  & 11.22& 14.1    & 19.4 & $<$2.40    \\ 
J1110+0321C  & 7.6    &   1.0 &$<$0.132 &$<$2.80  & 11.74&18.8    & 11.2 & $<$0.85    \\ 
J1110+0321E  & 155    &   1.1 &$<$0.007 &$<$0.13  & .... & 37.9    & 22.5 &    $-$     \\ 
J1110+0321W  & 223    &   0.9 &$<$0.004 &$<$0.08  &....&25.7    & 15.3 &    $-$     \\ 
J1228+3706   & 298    &   1.1 &$<$0.004 &$<$0.07  & 3.38 & 6.2     & 15.0 &  0.49$\pm$0.13 \\ 
J1241+6332C  & 67.9   &   0.7 &   0.157 &~~~2.90  & 4.42 & 4.4     & 11.0 &  1.01$\pm$0.11 \\ 
J1241+6332E  & 6.21   &   0.7 &$<$0.113 &$<$2.50  &....&13.6    & 34.0 &    $-$     \\ 
J1241+6332W  & 22.9   &   0.7 &$<$0.031 &$<$0.62  &....&21.2    & 53.0 &    $-$     \\ 
\hline
\end{tabular}
\end{center}
\begin{flushleft}
Column 1: Radio component. Column 2: Peak flux density in mJy.
Column 3: Spectral rms in mJy\,beam$^{-1}$\,channel$^{-1}$. 
Column 4: Maximum of optical depth or 1$\sigma$ limit to it.
Column 5: Integrated 21-cm optical depth or 3$\sigma$ upper limit in case
of non-detections with data smoothed to 10\,km\,s$^{-1}$. 
Column 6: Petrosian radius that contains 90\% of the Petrosian flux from galaxy 
as measured using $r$-band SDSS images.
Columns 7 and 8: Angular separation and the projected distance between the 
radio component (quasar) and centre of galaxy respectively.
Column 9: Rest equivalent width of \caiia~ or 3$\sigma$ upper limit to it. \\
\end{flushleft}
\label{gmrtres}
\end{table*}

\subsection{Details of the QGP with 21-cm absorption detection}

\begin{figure}
\centerline{{
\psfig{figure=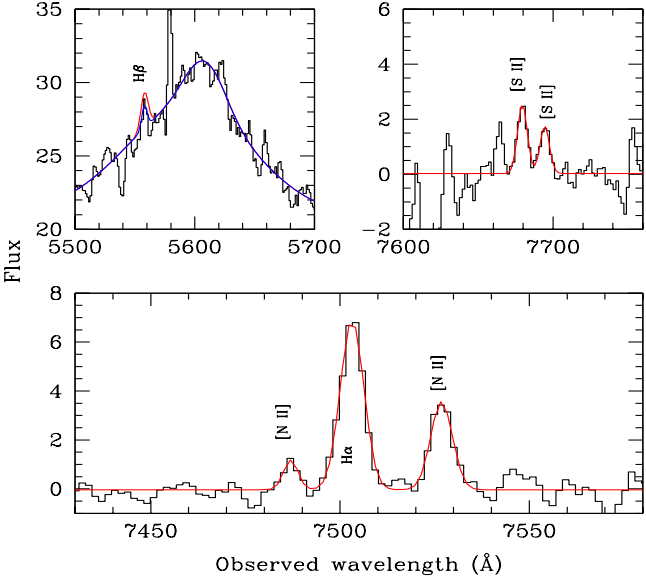,height=8.0cm,width=9.0cm,angle=0}
}}
\caption[]{Gaussian fits to the emission lines detected from 
the galaxy SDSS\,J124157.26+633237.6. The H$\beta$ line falls on
the C~{\sc iv} emission line of the QSO. Double Gaussian
fit to this broad emission line is also shown.
}
\label{eline}
\end{figure}

%
{\bf Quasar ($z_q$ = 2.625 SDSS\,J124157.54+633241.6) $-$ 
galaxy ($z_g$ = 0.143 SDSS\,J124157.26+633237.6) pair:}
The background quasar J124157.54+633241.6 is only at a projected 
separation of 4.4$^{\prime\prime}$ from the foreground galaxy at the redshift,
$z_g$ = 0.143. The quasar sight line is within the 
$r$-band Petrosian radius (R$_{90}$) that contains 90\% of the Petrosian
flux from galaxy (see Table~\ref{gmrtres}).
As can be seen from the Fig.~\ref{j1241sdsscol},  
the quasar sight line pierces through the stellar disk of 
the foreground galaxy. 

The galaxy redshift is measured using its emission lines superimposed
on top of the QSO continuum in the SDSS fiber spectrum. H$\alpha$,
H$\beta$, [S~{\sc ii}]$\lambda\lambda$6718,6732 and [N~{\sc ii}]$\lambda\lambda$6549,6585 are clearly detected.
These emission lines were fitted with Gaussians using $\chi^2$ 
minimization keeping the same redshift for all the lines (see Fig~\ref{eline}). 
The best fitted emission redshift is \zem = 0.1430$\pm$0.0001. 
The integrated emission line fluxes are summarized in Table~\ref{eline_t}.
The measured ratio of different line fluxes with respect to that of
H$\alpha$ in this system and from the SDSS template spectra of galaxies
are also given in this table. Clearly the line ratios that are indicators
of metallicity and reddening are consistent with average late type 
galaxies found in the SDSS.

\begin{figure}
\centerline{{
\psfig{figure=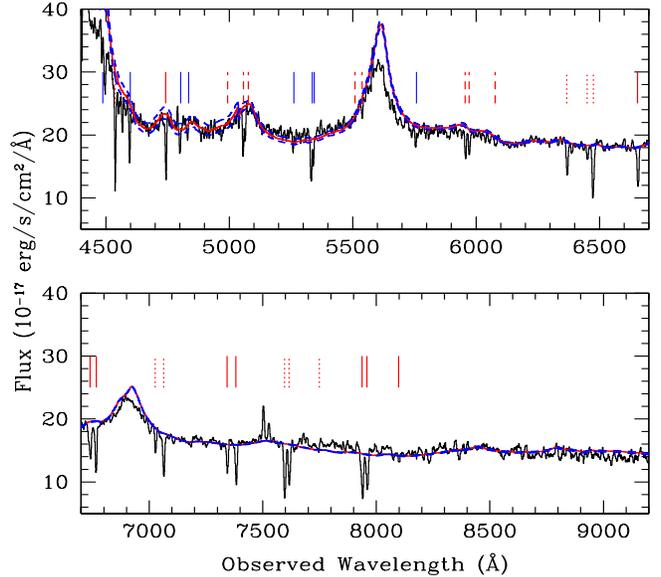,height=8.0cm,width=9.0cm,angle=0}
}}
\caption[]{SDSS spectrum of the QSO SDSS\,J124157.54+633241.6. The 
spectral energy distribution from the SDSS composite spectrum reddened
by the dust in the 21-cm absorbing galaxy is also over-plotted with the
associated 1$\sigma$ error. We also mark the locations of different
absorption lines originating from other intervening Mg~{\sc ii}
and C~{\sc iv} absorbers along the line of sight.  The doublet near 
7950\,\AA\, corresponds to \mgiiab~ of an absorbing system at $z_{abs}=1.840$ (solid red indicators).  
The doublets near 7600\,\AA\, and 5950\,\AA\, are \mgiiab\, from systems at $z_{abs}=1.717$ (dashed red indicators) 
and $z_{abs}=1.130$ (long dashed red indicators) respectively.  
The C~{\sc iv} absorber at $z_{abs}=2.444$ is marked with solid blue indicators. 
}
\label{sed}
\end{figure}

\begin{figure}
\centerline{{
\psfig{figure=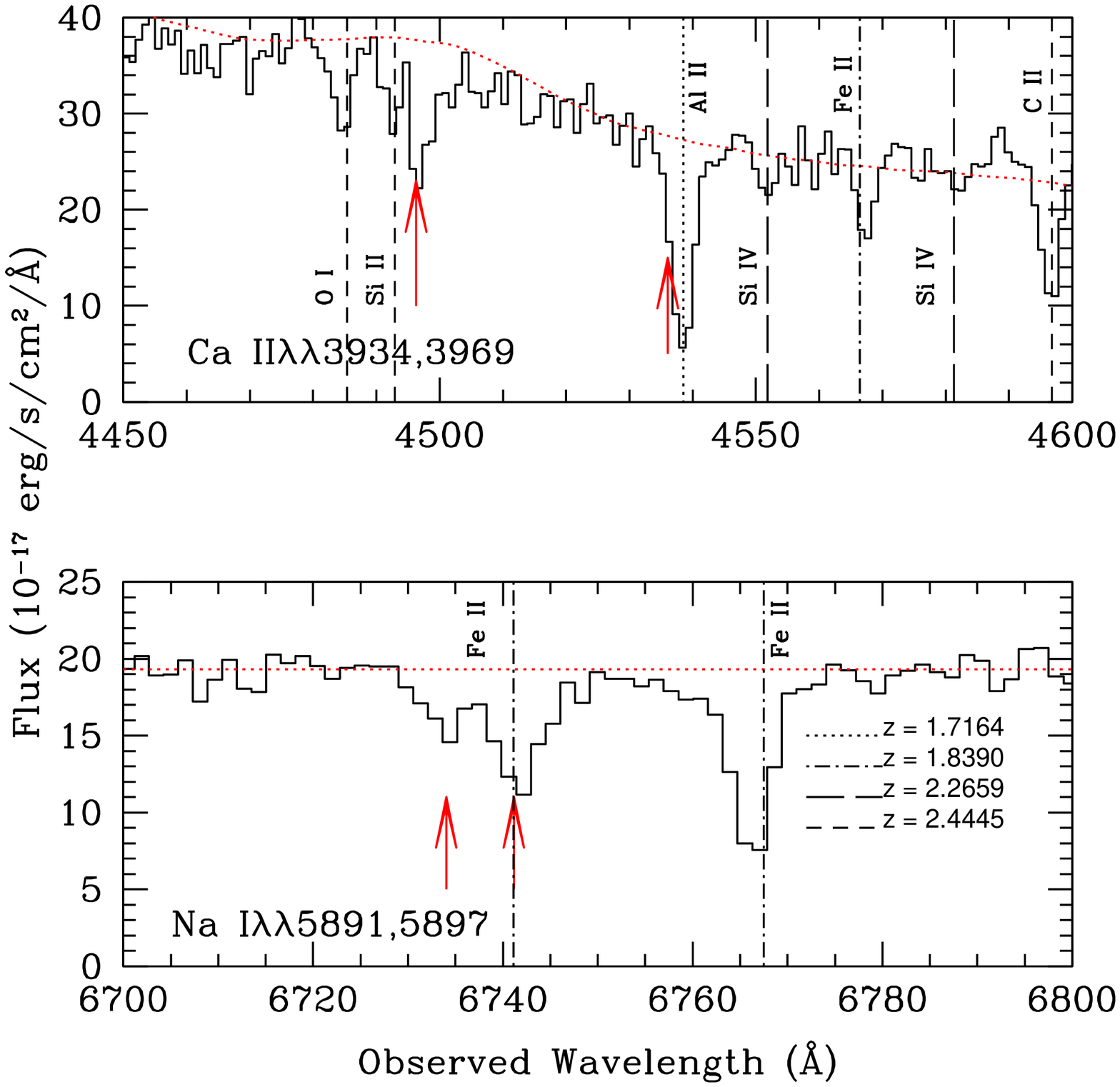,height=8.0cm,width=9.0cm,angle=0}
}}
\caption{SDSS spectrum of the QSO SDSS\,J124157.54+633241.6.
Arrows mark the expected locations of Ca~{\sc ii} and Na~{\sc i}
absorption from $z_g$ = 0.143 galaxy.  The vertical lines
mark the metal absorption lines from other intervening systems.
}
\label{ca2na1abs}
\end{figure}

%
The Balmer line ratio, F(H$\alpha$)/F(H$\beta$), has large error mainly
due to the fact that H$\beta$ is weak and present on top of the C~{\sc iv}
emission line of the QSO. Following Argence \& Lamareille (2009) we 
get the optical depth at the intrinsic V-band of the galaxy, 
$\tau_V^{\rm Balmer}$ = 1.4$^{+2.0}_{-1.3}$, using
\begin{equation}
{\rm
\tau_V^{Balmer} = {ln({H\beta \over H\alpha})- ln({H\beta^i \over H\alpha^i})
\over {\tau_\beta\over \tau_V}-{\tau_\alpha\over \tau_V}
}}.
\end{equation}
and the intrinsic Balmer ratio ${H\beta^i / H\alpha^i} = 2.85$ (Osterbrock
\& Ferland 2006) and $\tau_\lambda$ as given by the Eq. 3 of
Wild et al. (2007a). Such high
values are usually seen in high metallicity galaxies (see Fig. 1 of
Argence \& Lamareille 2009).

We get an independent estimate of $\tau_V$ and
E(B-V) along the sight line to the QSO by fitting the 
QSO spectral energy distribution (SED) using the SDSS
composite QSO spectra reddened by the Milky Way extinction curve 
and the method used in Srianand et al (2008)
and Noterdaeme et al. (2009). 
The best fit (with $\chi^2_\nu = 1.4$)
to the observed spectrum
(see Fig.~\ref{sed}) is obtained for A$_{\rm V}$= 0.44$\pm$0.04, 
where A$_\lambda$ = 1.086 $\tau_\lambda$.
Note that in principle the two measurements of $\tau_{\rm V}$
(or A$_{\rm V}$)
need not agree with the one 
based on Balmer decrement as the properties of the absorbing
gas along the QSO sight line need not be identical to that of the emitting 
region covered by the fiber.
If we use the 
relationship between  A$_{\rm V}$ and $N$(H~{\sc i}) observed for
the Milky Way we get
\begin{equation}
N(H~I) = (\frac{1}{\kappa})~7\times 10^{20} cm^{-2}.
\end{equation}
Here, $\kappa$ is the ratio of dust-to-gas ratio in the absorption system to that in the Milky Way.
Note that the extinction we find is higher than what is typically found
along the QSO sight lines with \mgii and \caii absorbers
(York et al. 2006; Wild, Hewett \& Pettini 2006) and less than that
seen towards absorption systems with 2175\,\AA\, dust absorption (Srianand et al. 2008). 
%
\begin{figure*}
\centerline{{
\includegraphics[trim= 13.40mm 0mm 0 0mm, clip, scale=0.70,angle=-90.0]{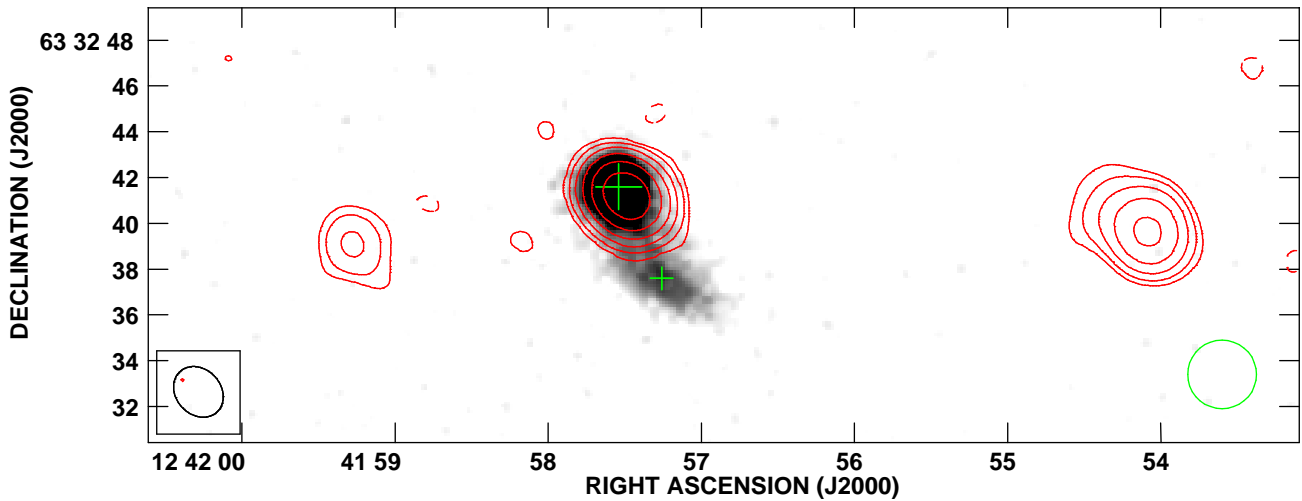}
}}
\caption[]{GMRT (contours) image overlaid on the SDSS $r$-band image of the
quasar ($z_q = 2.625$ SDSS J124157.54+633241.6)$-$galaxy 
($z_g$= 0.143  SDSS J124157.26+633237.6) pair.
The radio image has an rms of 0.4\,mJy\,beam$^{-1}$ and the restoring beam of
2.4$^{\prime\prime}\times$2.0$^{\prime\prime}$ with the P.A.=42$^\circ$.
The contour levels are 1.3\,$\times$\,(-1,1,2,4,8,16,32)\,mJy\,beam$^{-1}$.
The restoring beam for the radio image is shown as an ellipse and the optical positions
of the quasar and the galaxy are marked with a big and a small cross respectively. 
The extent of the SDSS spectroscopic fiber is shown at the bottom right corner of the image.
}
\label{j1241map}
\end{figure*}

%
Next we estimate the total star formation rate (SFR) in the portion of the
21-cm absorbing galaxy that is covered by the SDSS fiber. For this
we use (Argence \& Lamareille 2009) ,
\begin{equation}
log(SFR) = 0.95* log L(H\alpha) - log \eta_{H\alpha}.
\end{equation}
Adopting standard cosmological parameters, we 
get uncorrected luminosity of L(H$\alpha$) = 3.0 $\times 10^{40}$ erg/s. 
Using log ($\eta_{H\alpha}$) =  39.38 (Argence \& Lamareille 2009) we get  
SFR uncorrected for dust attenuation (and fiber filling factor) of 0.1 M$_\odot$ yr$^{-1}$. 
Using the integrated fluxes of
H$\alpha$ and N[~{\sc ii}]$\lambda$6549 lines and the linear relationship
given in Pettini \& Pagel (2004) we get 12+log(O/H) = 8.7, suggesting
super-Solar metallicity in the emission line region. This is 
slightly higher than the mean value measured by Argence \& Lamareille (2009)
for their whole sample. The inferred (O/H) and  $\tau_V^{\rm Balmer}$
values are consistent with those found for their high metallicity 
sub-sample. Interestingly, the inferred metallicity in this system
is higher than the values obtained by Noterdaeme, Srianand \& Mohan (2010)
for the sample of [O~{\sc iii}] emitting Mg~{\sc ii} absorbers at 
0.4$\le z\le 0.7$ and consistent with that measured in  a few strong
Ca~{\sc ii} absorbers (see Zych et al. 2007).

%
The SDSS spectrum of the quasar shows the Na~{\sc i}$\,\lambda$5891 and
Ca~{\sc ii}$\,\lambda$3934 absorption lines at the redshifted wavelengths
expected for the foreground galaxy (see Fig.~\ref{ca2na1abs}).
The other members of the Na~{\sc i} and Ca~{\sc ii} doublet are blended 
with the other intervening absorption lines. The rest equivalent
widths of Ca~{\sc ii}$\,\lambda$3934  and  Na~{\sc i}$\,\lambda$5891
absorption lines are 1.01$\pm$0.11 {\AA} and 0.85$\pm$0.10 {\AA} respectively. The 
measured rest equivalent width of Ca~{\sc ii}$\lambda$3934 is 
much higher than that measured by Nestor et al. (2008) in their
sample of DLAs at 0.6$\le z \le $1.3 and consistent with some of
the rare strong Ca~{\sc ii} absorbers in the sample of Wild
\& Hewett (2005).
%
\begin{table}
\caption{Emission line parameters of the galaxy SDSS\,J124157.26+633237.6}
\begin{center}
\begin{tabular}{l c c c}
\hline
\hline
 Line($l$) & F$_\lambda$(10$^{-17}$erg s$^{-1}$ cm$^{-2}$)&F(H$\alpha$)/F$_l$ &F$^c$(H$\alpha$)/F$^c_l$\\
\hline
H$\alpha$    & 56.6$\pm$4.3 & 1.0   & 1.0\\
H$\beta$     & 12.5$\pm$5.5 & 4.5$^{+4.2}_{-1.6}$& 2.8\\
\mbox{[N~{\sc ii}]}$\lambda$6549 & 10.0$\pm$3.5 &5.7$^{+3.7}_{-1.8}$ & 5.7\\ 
\mbox{[N~{\sc ii}]}$\lambda$6585 & 28.0$\pm$4.3 &2.0$^{+0.6}_{-0.4}$ & 2.0\\ 
\mbox{[S~{\sc ii}]}$\lambda$6718 & 12.9$\pm$3.6 &4.4$^{+2.1}_{-1.2}$ & 4.5\\ 
\mbox{[S~{\sc ii}]}$\lambda$6732 &  6.5$\pm$3.6 &8.7$^{+12.3}_{-3.5}$& 8.9\\ 

\hline
\end{tabular}
\end{center}
\begin{flushleft}
F$^C$: Flux corrected for dust attenuation.
\end{flushleft}
\label{eline_t}
\end{table}

\begin{figure}
\centerline{{
\psfig{figure=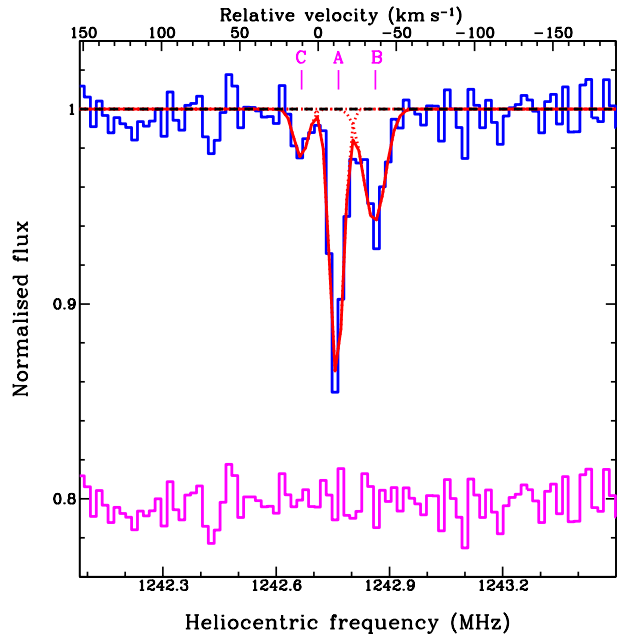,height=8.5cm,width=8.5cm,angle=0}
}}
\caption[]{GMRT spectrum of the 21-cm absorption detected towards the
quasar ($z_q = 2.625$ SDSS J124157.54+633241.6)$-$galaxy ($z_g$= 0.143 SDSS J124157.26+633237.6) pair.
Zero of the velocity scale is defined at $z_{em}$=0.14299 determined from the nebular emission 
detected in the SDSS spectrum.
Individual Gaussian components and the resultant fits to the absorption profile are plotted as
dotted and continuous lines respectively.
Residuals, on an offset arbitrarily shifted for clarity, are also shown.
}
\label{j1241abs}
\end{figure}

We detected 21-cm absorption at the expected redshifted frequency 
towards this QGP from the data obtained on 2008\,March\,03.  
In order to obtain better S/N and rule out the possibility of weak 
RFI mimicking the absorption feature, the object was reobserved 
on  2009\,June\,13 and 2009\,June\,14.  
These repeat observations resulted in the absorption feature consistent 
with the spectrum from the earlier observations and the velocity 
shift expected from the heliocentric motion of the earth.  
The GMRT image overlaid on the SDSS $r$-band image and the 
21-cm absorption spectrum towards the quasar are presented in  
Figs.~\ref{j1241map} and \ref{j1241abs} respectively.
The spectrum presented in Fig.~\ref{j1241abs} is obtained after 
combining the different epoch spectra.  
The absorption profile is well modelled with the three distinct 
Gaussian components A, B and C (Fig.~\ref{j1241abs}). 
The velocity offset of $\sim$10\,km/s between the strongest 21-cm absorption component 
and the redshift estimated from emission lines is within the redshift-measurement error.  
For a Gaussian profile, $N$(H~{\sc i}) and the peak optical depth ($\tau_{\rm p}$) are related by,
\begin{equation}
N({\rm HI}) = 1.93 \times 10^{18}~\tau_{\rm p}~\frac{T_{\rm s}}{f_{\rm c}}~\Delta v~{\rm cm}^{-2}.
\end{equation}
Here, $\tau_{\rm p}$ , $\Delta v$ and $f_{\rm c}$ are the peak optical depth, 
the FWHM in \kms of the fitted Gaussian, and the fraction of absorbing cloud covered by the 
radio source respectively. Individual
Gaussian fits and fit parameters are presented in Table~\ref{j1241gauss}.
The integrated 21-cm optical depth yields $N$(H~{\sc i})=5.3$\times10^{18}$T$_{\rm S}$\,cm$^{-2}$.  
If T$_{\rm S}$ = 100 K, as usually seen in the case of cold neutral medium,
then $N$(H~{\sc i}) will be 5$\times10^{20}$ cm$^{-2}$. This is very close to
the value we get from SED fitting using the Milky way extinction curve. 

In the GMRT image there are two faint radio components, one to the 
east (referred to as J1241+6332E) and one to the west (referred to as J1241+6332W)
of the quasar (Fig.~\ref{j1241map}).  It is not obvious if these radio components 
are associated with the quasar.  No 21-cm absorption was detected towards 
J1241+6332E and J1241+6332W (see bottom panels of Fig.~\ref{qgpnondet}) at an angular 
separation of 13.6$^{\prime\prime}$ and 21.2$^{\prime\prime}$ respectively from the 
galaxy (Table~\ref{gmrtres}).   
The flux density of these radio components is low and this prevents us from
getting any useful constraint on the 21-cm optical depth towards these sight lines.
%
%
%

%
\begin{table}
\caption{Details of the multiple Gaussian fits to the 21-cm absorption detected towards the
quasar ($z_q = 2.625$ SDSS J124157.54+633241.6)$-$galaxy ($z_g$= 0.143  SDSS J124157.26+633237.6) pair.
}
\begin{tabular}{lcccr}
\hline
\hline
Component$^\dag$ & $z_{abs}$ & $\Delta v^a$ &  $\tau^b_p$ &  {$f_c N(H\,I)^c \over T_s $}  \\
\hline
A    &   0.14294    &  10$\pm$1    & 0.148$\pm$0.010    &  0.29$\pm$0.03 \\
B    &   0.14285    &  16$\pm$2    & 0.060$\pm$0.007    &  0.19$\pm$0.03 \\
C    &   0.14303    &  11$\pm$4    & 0.025$\pm$0.009    &  0.05$\pm$0.02 \\
\hline
\multicolumn{5}{l}{$^\dag$ components as indicated in Fig.~\ref{j1241abs}}\\
\multicolumn{5}{l}{$^a$ FWHM in km s$^{-1}$; $^b$ peak optical depth}\\
\multicolumn{5}{l}{$^c$ in units of 10$^{19}$ cm$^{-2}$}\\
\end{tabular}
\label{j1241gauss}
\end{table}

%
\subsection{Details of QGPs with 21-cm absorption non-detections}
\begin{figure*}
\centerline{{
\includegraphics[scale=0.8]{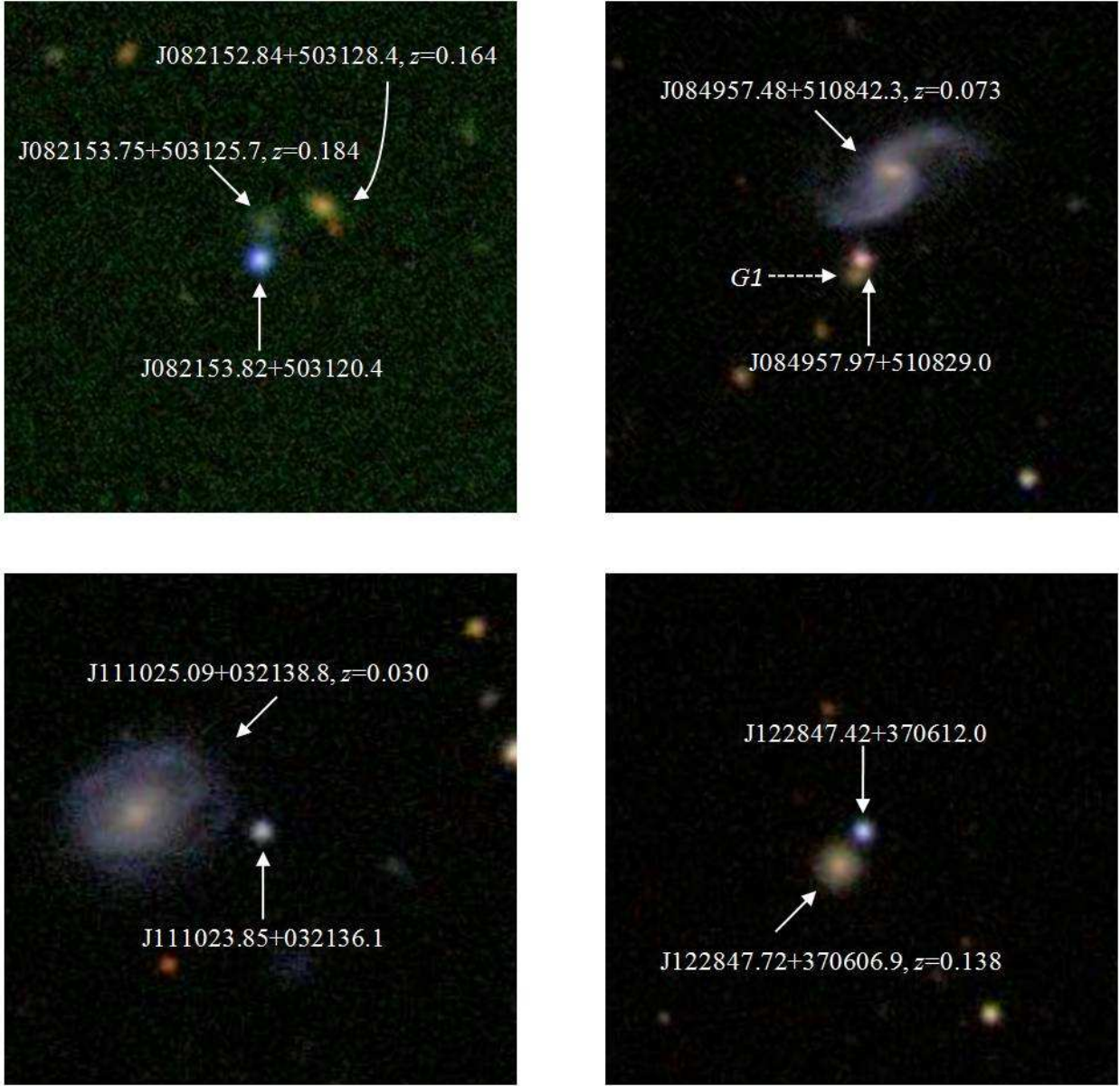}
}}
\caption[]{Color representations of the QGPs with 
no 21-cm absorption detections (see Section~3.2). 
Each image is 77$^{\prime\prime}$ on a side, with north-east towards the
top left corner, and centered on the QSO. Galaxies with known
redshifts are identified, along with their corresponding
redshifts. In the field of QSO J084957.97+510829.0 (top right panel), the object
labeled ``G1'' is the galaxy we suggest is responsible for the Ca~{\sc ii}
and Na~{\sc ii} absorption at $z_g=0.3120$, discussed in Section~3.2.2. 
}
\label{qgpsdsscol}
\end{figure*}

\begin{figure*}
\centerline{{
\psfig{figure=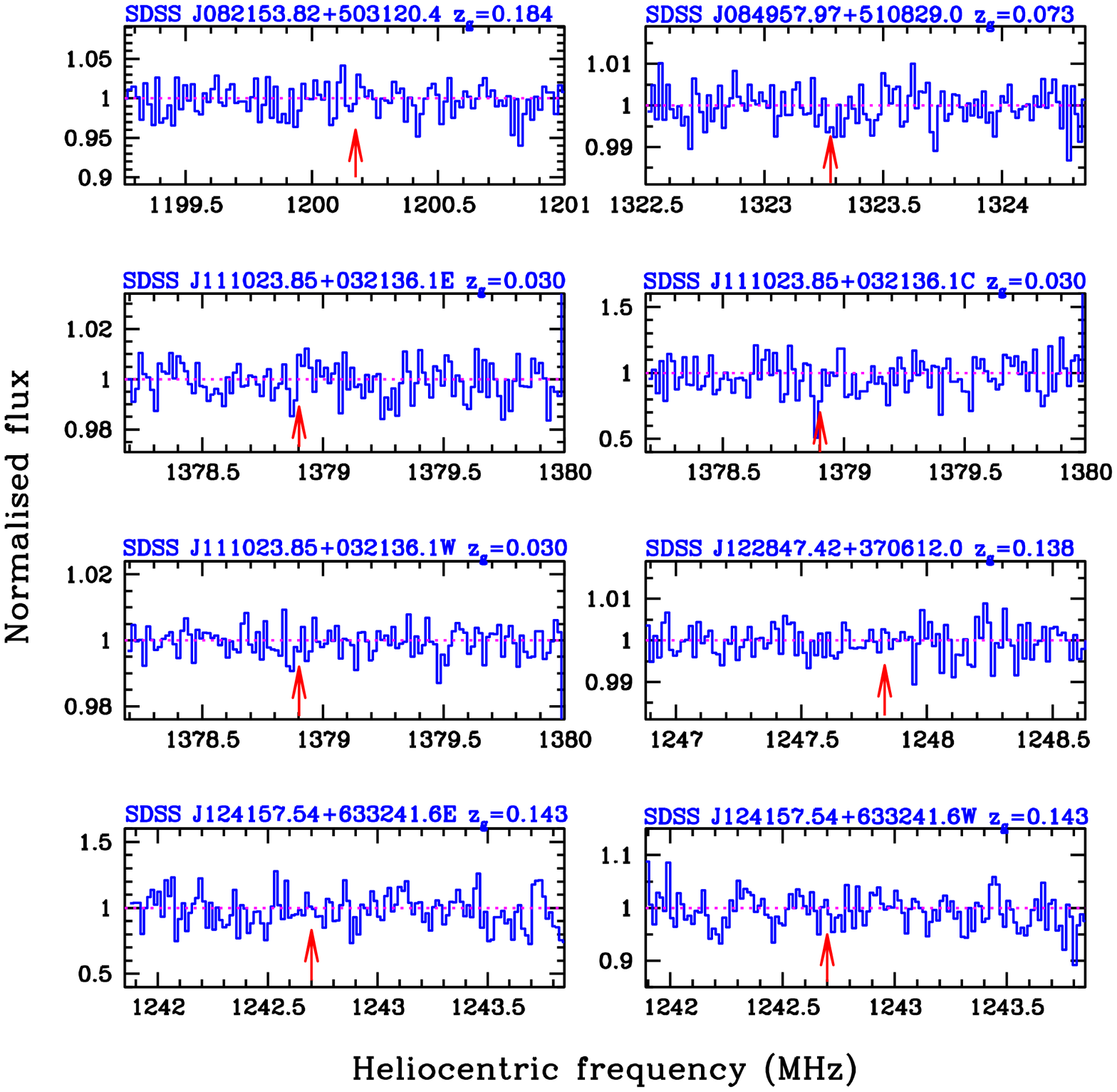,height=13.0cm,width=18.0cm,angle=0}
}}
\caption[]{GMRT spectra of QGPs with 21-cm absorption non-detections. 
Arrows mark the expected position of
21-cm absorption lines based on the galaxy redshift. 
}
\label{qgpnondet}
\end{figure*}

%
Here we discuss the QGPs that do not
show any detectable 21-cm absorption. The color representations of these 
4 QGPs are presented in Fig.~\ref{qgpsdsscol}.  The lack of 21-cm absorption
could be either related to low column density of H~{\sc i}
along the sight line and/or high spin-temperature, T$_{\rm S}$.
%
%
%
If a sight line passes through either the optical disk or 
an extended H~{\sc i} 21-cm line emitting region then one is sure of having
sufficient H~{\sc i} column density. For our sample we shall use
Petrosian radius (R$_{90}$) containing 90\% of the Petrosian flux 
measured using $r$-band SDSS images as 
an indicator of optical extent. These values are given in 
Table~\ref{gmrtres} for all the galaxies in our sample.
The extent of the H~{\sc i} gas in these galaxies is not known.
So we use the existing correlations between optical properties
and H~{\sc i} radius, (R$_{\rm H I}$), found
for the local galaxies. 
In the local universe, 
R$_{25}$ (the distance along the semi-major axis where the
B-band surface brightness falls to 25 mag\,arcsec$^{-2}$) is used
to quantify the optical radius of the galaxies.
It is also known that the extent of H~{\sc i} gas 
as measured from the 21-cm emission maps with 1$\sigma$ 
column density of few times 10$^{19}$\,cm$^{-2}$
is roughly 1.7 times R$_{25}$
(Noordermeer et al. 2005). As the scaling is found not
to depend strongly  on the  galaxy morphology (see Fig.~3
of Broeils \& Rhee 1997) we use this for all the galaxies.
Whenever possible 
we perform isophotal analysis using ellipse fitting to get
R$_{25}$ and the extent of H~{\sc i} gas, R$_{\rm H~I}$. 

As the galaxies in our sample are at higher redshifts (i.e $z\ge 0.03$)
any weak dependence of R$_{\rm H I}$ on the absolute magnitude
should also be considered.
Therefore, we also obtain R$_{\rm H~I}$ 
using absolute B-band magnitude and Equation 2 of Lah et al. (2009) 
derived from various correlations found 
by Broeils \& Rhee (1997). We get the B-band magnitude
using $r$- and $g$-band magnitudes of the galaxy 
measured in the SDSS images and the filter transformation
equations given by 
Lupton (2005)\footnote{http://www.sdss.org/dr6/algorithms/sdssUBVRITransform.html}. 
The above two estimates of  R$_{\rm H~I}$  allow us to have a rough
idea of possible H~{\sc i} extent in these galaxies.
In what follows we discuss each QGP in detail.

\subsubsection{ Quasar($z_q$ = 2.130 SDSS\,J082153.82+503120.4) $-$ 
galaxy ($z_g$ = 0.184 SDSS J082153.75+503125.7) pair} 
\begin{figure*}
\centerline{\hbox{
{\psfig{figure=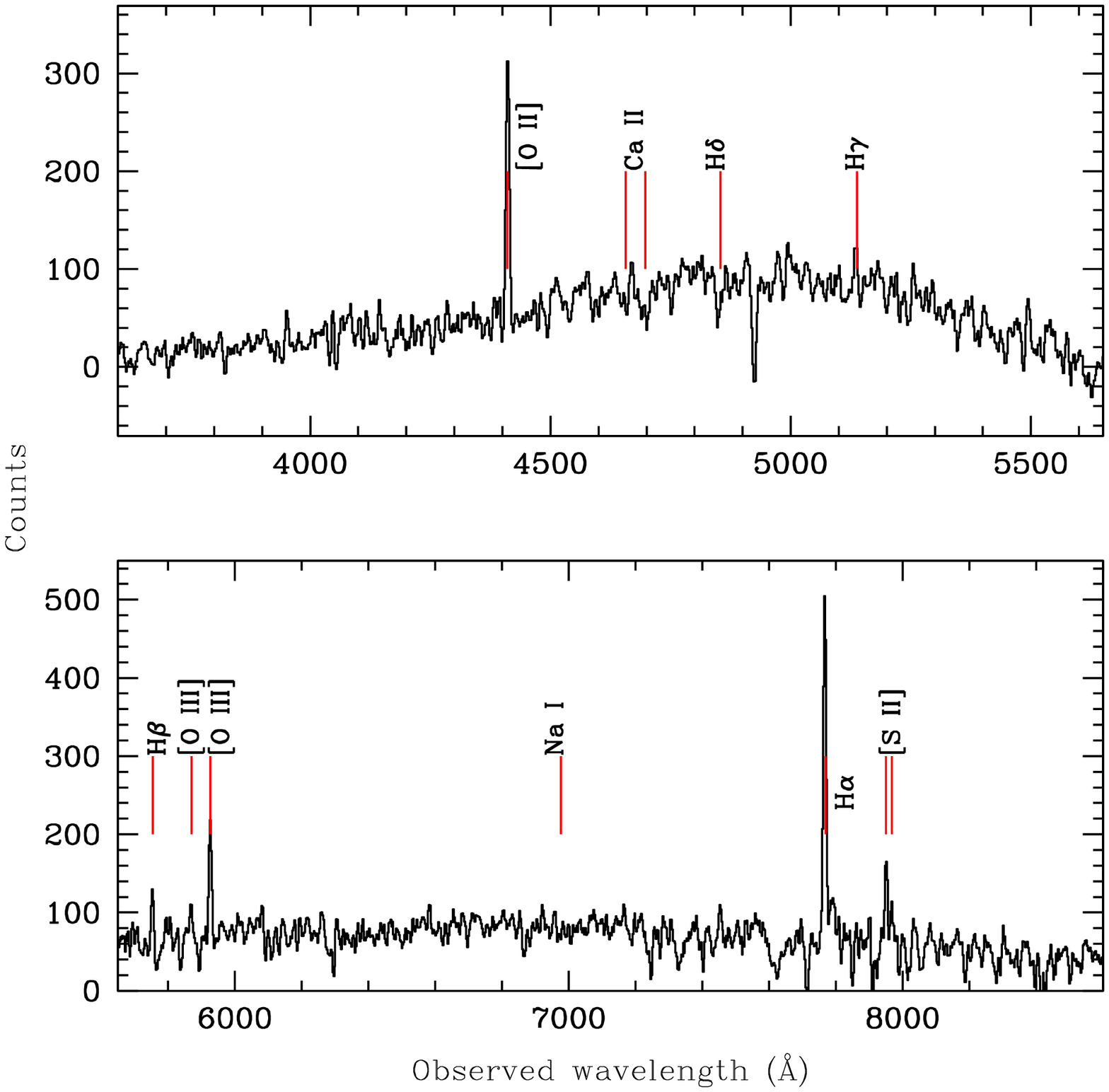,height=9.0cm,width=8.0cm,angle=0}}
{\psfig{figure=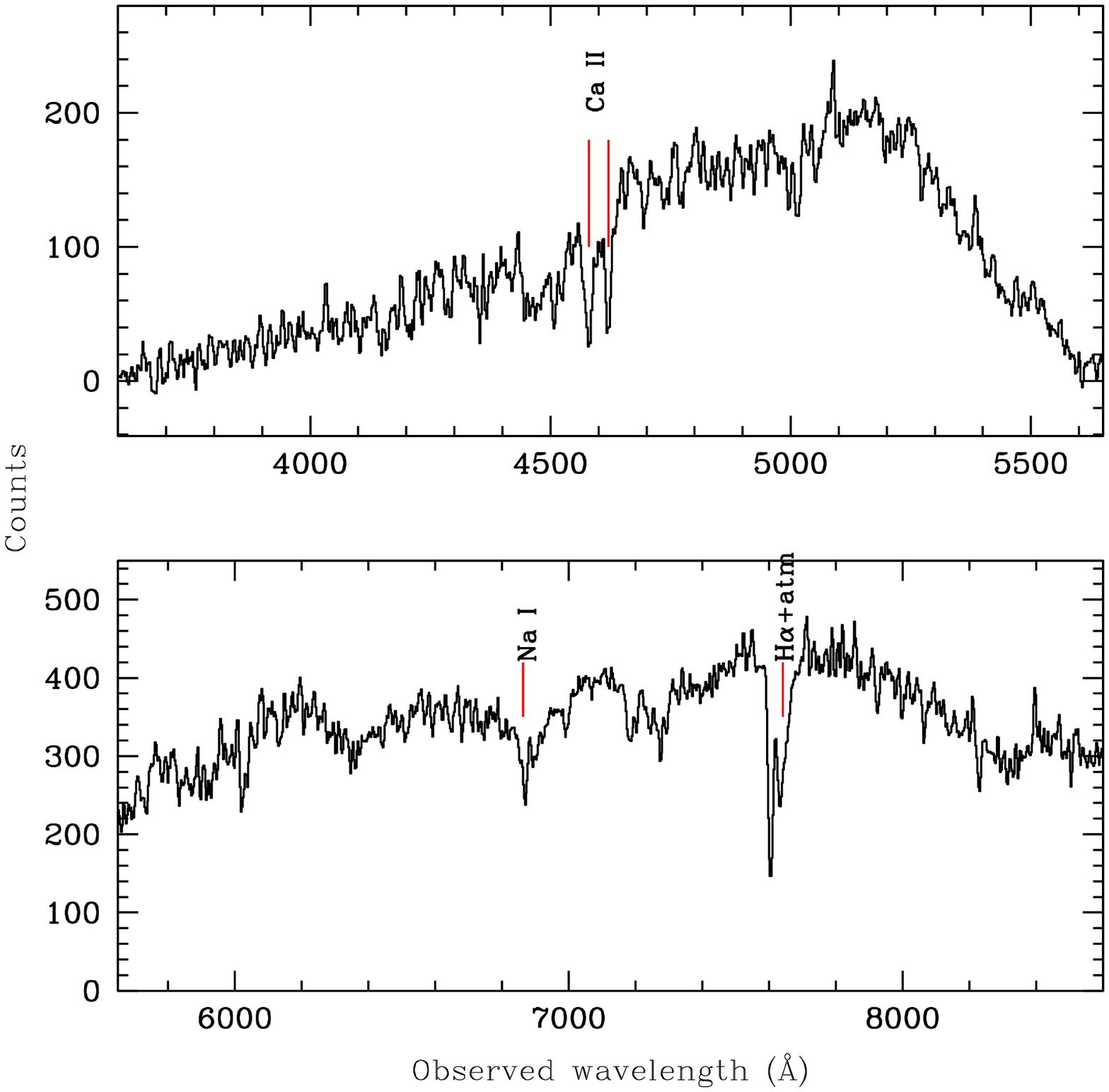,height=9.0cm,width=8.0cm,angle=0}}
}}
\caption[]{APO spectrum of the galaxy J082153.75+503125.7 at
$z_g$ = 0.1835 (left) and J082152.84+503128.4 at  $z_g$ = 0.1640 (right).}
\label{apo0821}
\end{figure*}
\begin{figure}
\centerline{{
\includegraphics[trim= 18.7mm 0mm 0 0mm, clip, scale=0.40,angle=-90.0]{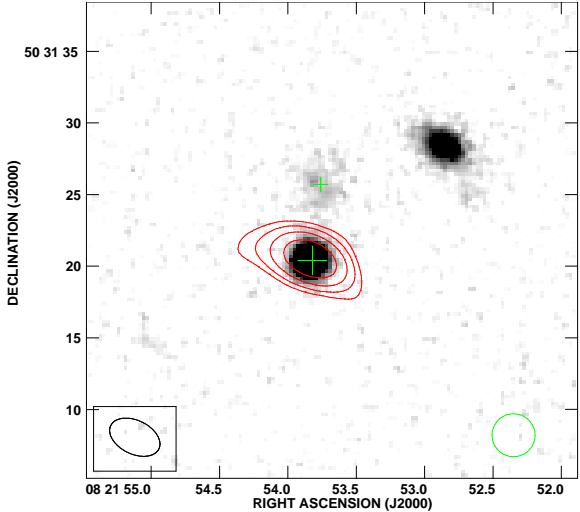}
}}
\caption[]{GMRT (contours) image overlaid on the SDSS $r$-band image of the
quasar($z_q$ = 2.130 SDSS\,J082153.82+503120.4) $-$ 
galaxy ($z_g$ = 0.184 SDSS J082153.75+503125.7) pair.
The radio image has an rms of 0.2\,mJy\,beam$^{-1}$ and the restoring beam of
3.7$^{\prime\prime}\times$2.3$^{\prime\prime}$ with the P.A.=64.0$^\circ$.
The contour levels are 2.1\,$\times$\,(-1,1,2,4,8)\,mJy\,beam$^{-1}$.
The restoring beam for the radio image is shown as an ellipse and the optical positions
of the quasar and the galaxies are marked with the big and small crosses respectively.
The extent of the SDSS spectroscopic fiber is shown at the bottom right corner of the image.
}
\label{j0821map}
\end{figure}

The background quasar J082153.82+503120.4 has a flux 
density of $\sim$54\,mJy in the 1.4\,GHz FIRST image and 
is the weakest background radio source in our sample of QGPs to 
search for 21-cm absorption.  
The SDSS $r$-band image shows that there are two galaxies within 
0.2$^{\prime}$ of the quasar. The nearest 
of these i.e. J082153.75+503125.7 is at a separation 
of 5.2$^{\prime\prime}$ whereas the other galaxy i.e. 
J082152.84+503128.4 is at the distance of 
12.2$^{\prime\prime}$ (see top left panel of Fig.~\ref{qgpsdsscol}).

We measured the redshift of the galaxy  
J082153.75+503125.7 using our APO spectrum to be $z_g$=0.1835. 
(Fig.~\ref{apo0821}). 
Data were taken on 2007\,March\,14 using DIS at APO, with
the B400/R300 grating, with a 1.5$^{\prime\prime}$ slit and 
the exposure time was 1200\,sec. 
While obtaining the spectrum of the closest galaxy
(J082153.75+503125.7), the DIS slit was
orientated to also cover the second galaxy i.e. 
J082152.84+503128.4. The spectrum of the second galaxy showed no emission
lines, but strong Ca~II H \& K, G-band, and Na~I stellar absorption
lines. To calculate the redshift of this galaxy, we used the IRAF routine
{\tt fxcor} to cross-correlate its spectrum with that of a Galactic
star with a known radial velocity, HD~182572. We derived a
redshift of $z=0.1640$, with an error of $\simeq 200$~\kms . In this
case, the error is dominated by the error given by {\tt fxcor}, which
is larger than the error from the wavelength calibration of the
spectrum itself.
We searched for 21-cm absorption only in the 
galaxy SDSS J082153.75+503125.7 at $z_g$=0.184 and do 
not discuss the second galaxy hereafter.

The galaxy J082153.75+503125.7 is faint (M$_B$ = -19.1 mag
and L/L$_B^*$ = 0.19 if we use M$_B^*$ = -20.90 as found by
Marinoni et al. 1999) and compact
with R$_{90}\sim3.5''$ with a corresponding physical
extent of 10.7 kpc.  We estimate the effective
H~{\sc i} radius R$_{\rm H~I}$ of 8.2 kpc from the B-band
absolute magnitude. As the 
object is faint we could not perform isophotal
analysis for this galaxy. 
However, it is clear that optical extent and estimated
H~{\sc i} radius are significantly smaller than the impact
parameter of the galaxy from the QSO sight line
(i.e 15.9 kpc).

In the SDSS QSO spectrum we do not
find Ca~{\sc ii} and Na~{\sc i} absorption at the $z_g$. 
The 3$\sigma$ upper limit for W$_{\rm r}$(\caiia) is 
0.25\,{\AA}. The spectral energy
distribution of the QSO is also consistent with very
little reddening. Based on the SED fits we find
$\tau_V$ = 0.10$\pm$0.02 and E(B-V) = 0.04$\pm$0.01.
The GMRT image overlaid on the SDSS $r$-band image is presented 
in Fig.~\ref{j0821map}.
No 21-cm absorption is detected towards this radio source at the 
impact parameter of 15.9\,kpc (Fig.~\ref{qgpnondet}).  
The 3$\sigma$ upper limit on the H~{\sc i} column density is 
$N$(H~{\sc i})=7.0$\times10^{17}{\rm T_S}$\,cm$^{-2}$. As the impact
parameter is larger than optical radius and the H~{\sc i}
radius estimated above the lack of absorption could be 
simply due to the lack of sufficient neutral gas along the line of sight.

\noindent
\subsubsection{Quasar($z_q$ = 0.584 SDSS\,J084957.97+510829.0) $-$ 
galaxy ($z_g$ = 0.073 SDSS\,J084957.48+510842.3) pair}
\begin{figure}
\centerline{{
\includegraphics[trim= 18.2mm 0mm 0 0mm, clip, scale=0.40,angle=-90.0]{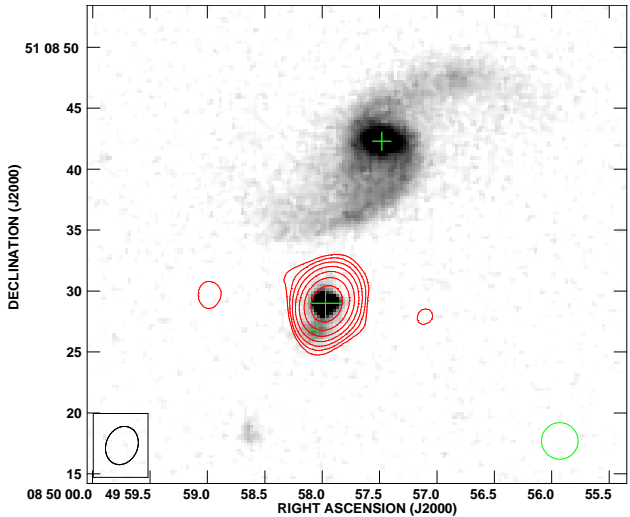}
}}
\caption[]{GMRT (contours) image overlaid on the SDSS $r$-band image of the
quasar($z_q$ = 0.584 SDSS\,J084957.97+510829.0) $-$ galaxy 
($z_g$ = 0.073 SDSS\,J084957.48+510842.3) pair.
The radio image has an rms of 0.4\,mJy\,beam$^{-1}$ and the restoring beam of
3.2$^{\prime\prime}\times$2.6$^{\prime\prime}$ with the P.A.=$-$22.6$^\circ$.
The contour levels are 2.1\,$\times$\,(-1,1,2,4,8,16,32,64,128)\,mJy\,beam$^{-1}$.
The restoring beam for the radio image is shown as an ellipse and the optical positions
of the quasar and the galaxy are marked with a big and a small cross respectively.
The extent of the SDSS spectroscopic fiber is shown at the bottom right corner of the image.
}
\label{j0849map}
\end{figure}

The background radio source is a well-known object which has a BL Lacertae spectrum 
during outbursts (Arp et al. 1979).  
The redshift of this object was measured to be $z_q$ = 1.860 on the basis of the two 
emission lines which were identified as C{\sc iv}$\lambda$1549 and 
C{\sc iii}$]\lambda$1909 (Arp et al. 1979).  
The actual redshift as estimated from the SDSS spectrum with much wider 
spectral coverage detecting the emission lines of Mg~{\sc ii}, O{\sc [iii]}, 
H$\beta$, H$\gamma$ and H$\delta$ has been found to be \zem=0.584.
The radio source is situated at 14.1$^{\prime\prime}$ south of the galaxy 
J084957.48+510842.3 at $z_g$=0.073 which is the southern member of the 
interacting pair of galaxies (Stickel, Fried \& K\"uhr 1989).  
This is the brightest galaxy in our sample (M$_B$ = -20.7 and L$_B$/L$_B^*$ = 0.8). 
Based on the absolute B-band magnitude we expect the H~{\sc i} emission
to extend up to a radius of 14 kpc.
We used the redshift of $z_g$=0.0734 (measured from the SDSS spectrum of the galaxy) 
to search for 21-cm absorption from 
the halo of foreground spiral galaxy.  The GMRT image overlaid on  
the SDSS $r$-band image is shown in Fig.~\ref{j0849map}.  The radio source 
is compact in our image and no 21-cm absorption was detected 
towards this sight line at $z_g=0.0734$ with the impact parameter of 19.4\,kpc 
(Fig.~\ref{qgpnondet}). The corresponding 3$\sigma$ upper limit on 
the H~{\sc i} column density is $N$(H~{\sc i})=1.5$\times$10$^{17}{\rm T_S}$\,cm$^{-2}$. 

This QGP has been previously searched for 21-cm absorption 
and molecular absorption by Boiss\'e et al. (1988) and Wiklind \& Combes (1995) 
respectively.  Boiss\'e et al. (1988) obtained a shallower spectrum of this 
radio source using the Nancay radio telescope yielding 3$\sigma$ 21-cm optical 
depth limit, $\int\tau_{3\sigma}dv<$0.32\,km\,s$^{-1}$ as compared to the 
limit of 0.08\,km\,s$^{-1}$ presented here. Absorption lines of Ca~{\sc ii}
and Na~{\sc i} are not detected in the SDSS QSO spectrum. However, since quasar 
is faint, the equivalent width upper limits are poor (Table~\ref{gmrtres}).
There are speculations about the optical spectrum of this blazar being affected by
foreground galaxy lensing and/or reddening
(Stickel et al. 1989; Ostman 2006). It is known that dusty absorbers
tend to show strong 21-cm absorption (cf. Srianand et al. 2008 and the 
detection presented in Section 3.1). The absence of 21-cm absorption
is thus inconsistent with the dust reddening observed in the spectrum of the QSO 
by the galaxy J084957.97+510829.0.
If we use the average spin temperature of 500\,K measured in the
QGPs (Carilli \& van Gorkom 1992), our observations are consistent
with $N$(H~{\sc i})$\le 7.5\times10^{19}$ cm$^{-2}$. This is too low 
to produce any detectable extinction even if the dust-to-gas ratio is
close to what we see in the Milky way. 

\begin{figure}
\centerline{{
\includegraphics[trim= 0mm 0mm 0 0mm, clip, scale=0.70,angle=000.0]{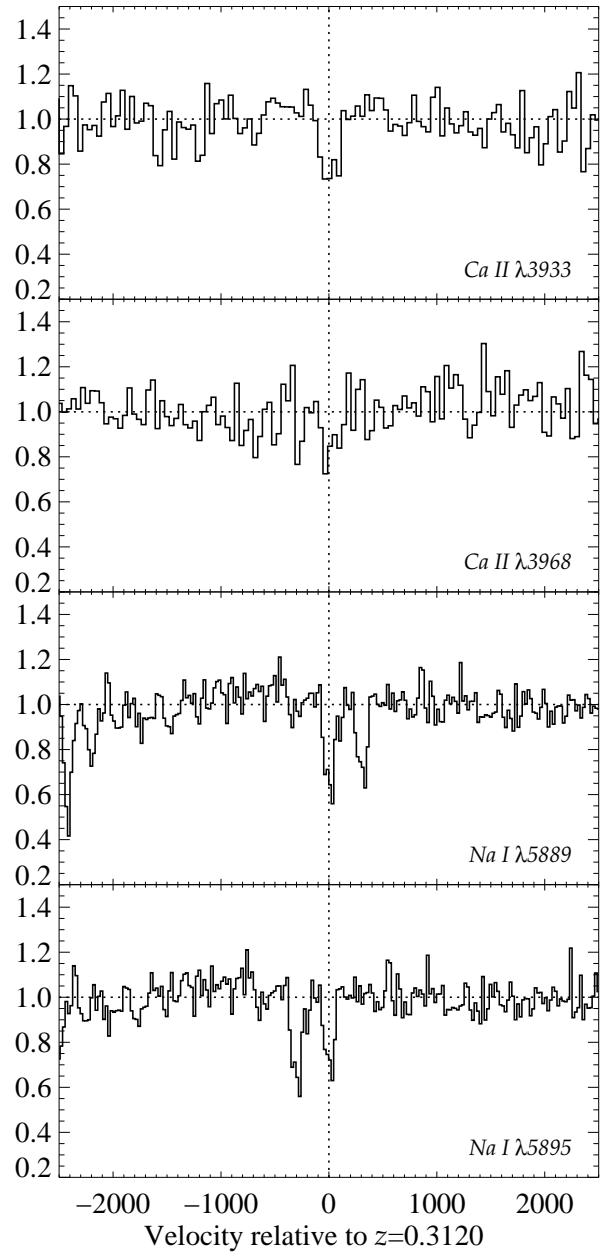}
}}
\caption[]{ Velocity plot for the \nai and \caii absorption lines (wavelengths are in 
air) detected towards the quasar, SDSS\,J084957.97+510829.0.
}
\label{j0849apospec}
\end{figure}

We have detected strong \nai and \caii absorption at $z_{abs}$=0.3120 towards this 
quasar sight line through our APO 3.5-metre observations on 2007\,May\,22 (Fig.~\ref{j0849apospec}).  
These observations used DIS R1200/B1200 gratings
with a 1.5$^{\prime\prime}$ slit and the exposure time was 3x1500\,sec.
Rest frame equivalent widths for \nai and \caii absorption lines are 
W$_{\rm r}$(\naia)=0.84$\pm$0.25,         W$_{\rm r}$(\naib)=0.60$\pm$0.22,      
W$_{\rm r}$(\caiia)=0.60$\pm$0.18  and    W$_{\rm r}$(\caiib)=0.39$\pm$0.16.      
It is possible that this strong \caii absorber originates from the 
{\sl galaxy} (labelled `G1') blended with quasar in the 
SDSS image (Fig.~\ref{qgpsdsscol}) and is responsible for 
the reddening of quasar. 
Unfortunately, the GMRT data to search for 21-cm absorption in this 
system were rendered unusable due to strong RFI.

%

%
\noindent
\subsubsection{Quasar($z_q$ = 0.966 SDSS\,J111023.85+032136.1) $-$ 
galaxy ($z_g$= 0.030 SDSS\,J111025.09+032138.8) pair}
\begin{figure*}
\centerline{{
\includegraphics[trim= 13.55mm 0 0 0, clip, scale=0.7,angle=-90.0]{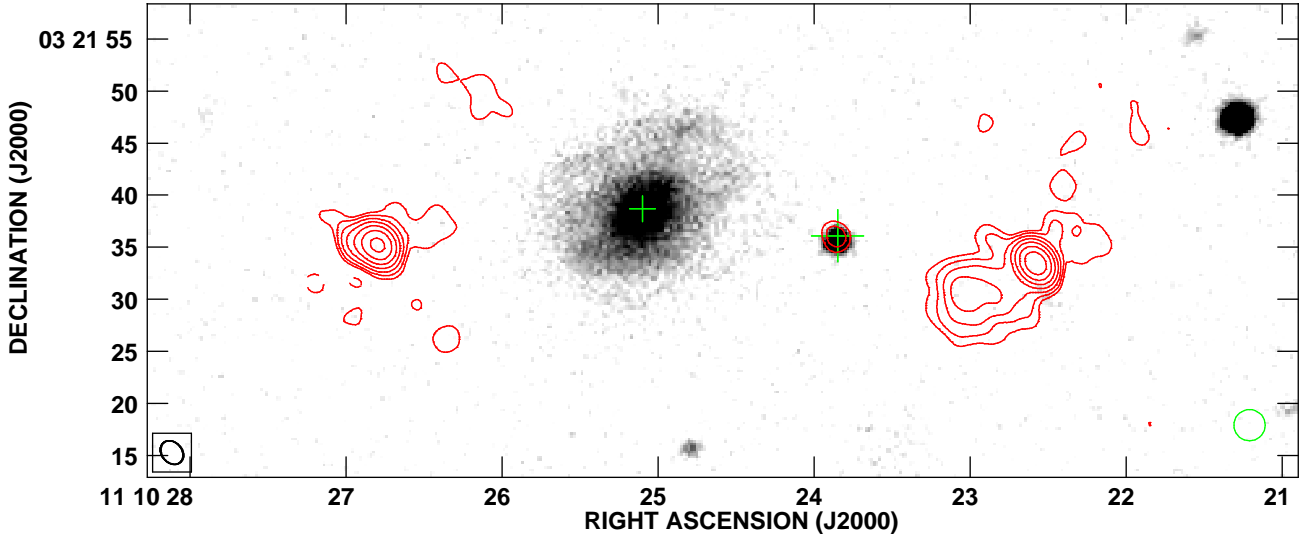}
}}
\caption[]{GMRT (contours) image overlaid on the SDSS $r$-band image of the
quasar($z_q$ = 0.966 SDSS\,J111023.85+032136.1) $-$ 
galaxy ($z_g$= 0.030 SDSS\,J111025.09+032138.8) pair.
The radio image has an rms of 0.4\,mJy\,beam$^{-1}$ and the restoring beam of
2.5$^{\prime\prime}\times$1.9$^{\prime\prime}$ with the P.A.=40$^\circ$.
The contour levels are 2.0\,$\times$\,(-1,1,2,4,8,16,32,64)\,mJy\,beam$^{-1}$.
The restoring beam for the radio image is shown as an ellipse and the optical positions
of the quasar and the galaxy are marked with a big and small a cross respectively.
The extent of the SDSS spectroscopic fiber is shown at the bottom right corner of the image.
}
\label{j1110map}
\end{figure*}

\begin{figure}
\centerline{{
\psfig{figure=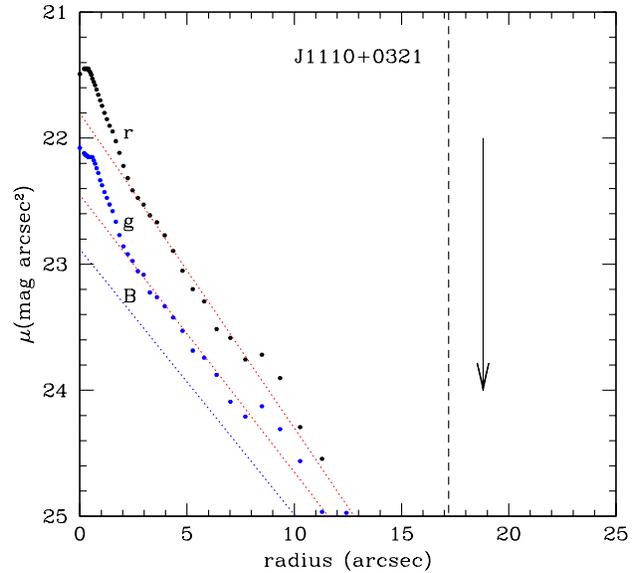,height=8.0cm,width=8.5cm,angle=0}
}}
\caption[]{The $r$- and $g$-band surface brightness profile of the galaxy 
J111025.09+032138.8 with best fitted exponential profiles over-plotted.
We construct the B-band surface brightness profile using 
B = $g$ + 0.3130($g-r$) + 0.2271 and estimate R$_{25}$. 
The vertical dashed line marks the expected radius of
the H~{\sc i} gas based on the relationship between H~{\sc i} radius 
and R$_{25}$ obtained by Noordermeer et al. (2005). The arrow marks
the impact parameter for the core of the QSO.
}
\label{sur1110}
\end{figure}

Murdoch et al. (1983) presented detailed spectroscopy of the 
background quasar as well as the foreground galaxy that has 
been classified as Sc.  
This is the lowest redshift galaxy in our sample.
The SDSS quasar spectrum does not show
Ca~{\sc ii} absorption lines at the redshift of the galaxy.  
The radio emission from the quasar is well resolved into 
a triple in our GMRT image where most of the emission is contained 
in the two lobes on either side of the foreground galaxy (Fig.~\ref{j1110map}).  
The radio lobe to the west of the foreground galaxy is at a projected
separation of $\sim$25.7$^{\prime\prime}$ (15.3\,kpc) whereas 
the radio lobe towards the east is at the angular separation of 
$\sim$37.9$^{\prime\prime}$ (22.5\,kpc).
The radio core is at the separation of 18.8$^{\prime\prime}$ 
from the galaxy but is unfortunately very weak with the peak 
flux density of only 7.6\,mJy\,beam$^{-1}$.
The radio source therefore provides multiple sight lines close 
to the foreground galaxy. The galaxy has Petrosian radius
R$_{90}\sim11.74''$. We use the $r$- and $g$-band SDSS images and
ellipse fitting routine in STSDAS package in IRAF to get the
surface brightness profile of the galaxy (see Fig~\ref{sur1110}).
We get R$_{25}$, the radius at which the B-band surface brightness
reaches 25 mag\,arcsec$^{-2}$, using these profiles and
B = $g$ + 0.3130($g-r$) + 0.2271.
This together with the well known 
relationship between the radius of the H~{\sc i} emitting
gas and R$_{25}$ measured for Sa/Sab galaxies (Noordermeer et al. 2005)
we expect the H~{\sc i} gas to extend upto 17$''$ (i.e $\sim 10$ kpc)
from the center of the galaxy. The absolute B-band magnitude of this 
galaxy is -18.6 mag and L$_B$/L$_B^*$ = 0.12. The R$_{\rm H~I}$ expected
from the absolute B-band magnitude is 6.8\,kpc. 
Therefore, if this galaxy follows standard scaling relation
assumed here we expect only the sight line towards the radio core 
to pass through the H~{\sc i} disk.
%

No 21-cm absorption was detected towards any of the radio components.  
The absorption spectra towards the peak components of the eastern and 
western radio lobes (referred to as J1110+0321E and J1110+0321W) 
and the radio core (i.e. J1110+0321C) are presented in the Fig.~\ref{qgpnondet} 
and the corresponding optical depth limits are provided in the Table~\ref{gmrtres}.
In the case of the core, the sight line may just graze through the 
outer H~{\sc i} disk.  But due to the weakness of radio emission, the 
21-cm absorption optical depth limit is very high. 
The integrated optical depths typically seen in the QGPs are much lower than this
limit (see Section~4.1). 
The best 3$\sigma$ upper limit on the H~{\sc i} column density, 
$N$(H~{\sc i})=1.5$\times10^{17}$T$_{\rm S}$\,cm$^{-2}$, is obtained towards the 
J1110+0321W component. 
%

%

\noindent
\subsubsection{ Quasar ($z_q$ = 1.517 SDSS\,J122847.42+370612.0) $-$ 
galaxy ($z_g$ = 0.138 SDSS\,J122847.72+370606.9 pair):}
\begin{figure}
\centerline{{
\includegraphics[trim= 18.8mm 0mm 0 0mm, clip, scale=0.40,angle=-90.0]{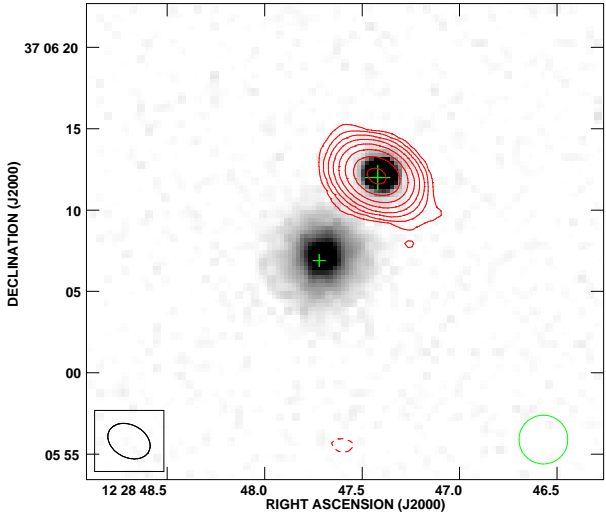}
}}
\caption[]{GMRT (contours) image overlaid on the SDSS $r$-band image of the
quasar ($z_q$ = 1.517 SDSS\,J122847.42+370612.0) $-$ 
galaxy ($z_g$= 0.138 SDSS J122847.72+370606.9) pair.
The radio image has an rms of 0.4\,mJy\,beam$^{-1}$ and the restoring beam of
2.7$^{\prime\prime}\times$2.0$^{\prime\prime}$ with the P.A.=62$^\circ$.
The contour levels are 2.1\,$\times$\,(-1,1,2,4,8,16,32,64,128)\,mJy\,beam$^{-1}$.
The restoring beam for the radio image is shown as an ellipse and the optical positions
of the quasar and the galaxy are marked with a big and a small cross respectively.
The extent of the SDSS spectroscopic fiber is shown at the bottom right corner of the image.
}
\label{j1228map}
\end{figure}

The foreground galaxy at the redshift of 0.1383 is at the projected 
separation of (6.2$^{\prime\prime}$) 15\,kpc (Fig.~\ref{j1228map}).
The galaxy in this case is face-on with R$_{90}$ = 4.4$^{\prime\prime}$. 
%
\begin{figure}
\centerline{{
\psfig{figure=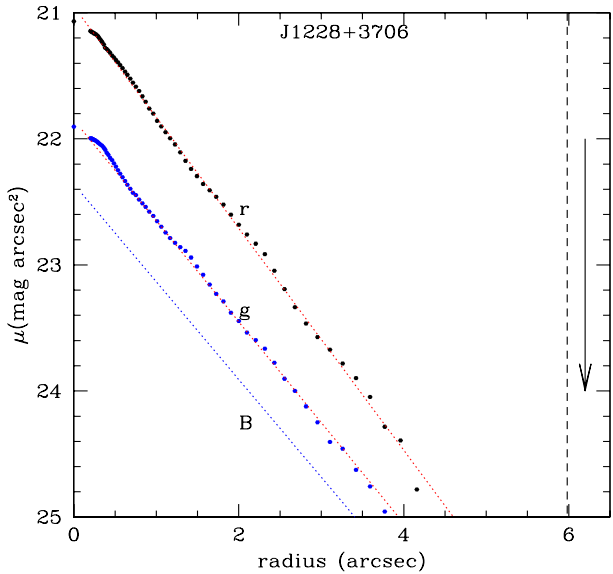,height=8.0cm,width=8.5cm,angle=0}
}}
\caption[]{The $r$- and $g$-band surface brightness profile of the galaxy 
J122847.72+370606.9 with best fitted exponential profiles over-plotted.
Rest are similar to that of Fig.~\ref{sur1110}.
}
\label{sur1228}
\end{figure}
%
We also performed the ellipse fitting to this galaxy and estimated
the possible extent (R$_{\rm H~I}$ = 14 kpc) of H~{\sc i} gas in 
this galaxy (Fig.~\ref{sur1228}). 
The absolute B-band magnitude of the galaxy is -19.9 mag and $L_B/L_B^*$
is 0.41. We derive R$_{\rm H~I}$ = 11 kpc from the absolute B-band
magnitude. This suggests that the QSO sight line may just graze through
the outer part of the extended H~{\sc i} gas.
The redshift of the galaxy is accurately measured from the available
SDSS spectrum of the galaxy. Weak absorption features with rest equivalent widths
W$_r$(\caiia)=0.49$\pm$0.13\,\AA\, and 
W$_r$(\caiib)=0.20$\pm$0.13\,{\AA}\, are 
detected at the expected positions of Ca~{\sc ii}
doublet in the SDSS QSO spectrum. However, Na~{\sc i} absorption
lines are not detected.
The GMRT image overlaid on the SDSS $r$-band image is presented 
in Fig.~\ref{j1228map}.
No 21-cm absorption is detected towards this QGP 
(Fig.~\ref{qgpnondet}) with the 3$\sigma$ upper limit on the 
H~{\sc i} column density being $N$(H~{\sc i})=1.3$\times10^{17}$T$_{\rm S}$\,cm$^{-2}$.

\section{Discussion}
In order to investigate the relationship between 21-cm absorption 
and the properties of the absorbing galaxy, we combine our results 
with 21-cm absorption measurements of all the QGPs available in 
the literature (Table~\ref{qgplit}).  
Observations of the QGP 3C268.4/NGC4138 available 
from Haschick, Crane \& Baan (1983) are not considered here.  
In this case, 21-cm absorption superimposed on the 21-cm emission is inferred from 
the sharp interruption in the emission profile obtained using Green 
Bank 300-ft antenna and the absorption optical depth is highly uncertain. 
In addition to this list, 21-cm 
absorption has been detected towards the radio source 3C178 
from the foreground galaxy NGC2377 (Haschick et al. 1980).  
The nature of background source is not clear in this case.  Radio 
source is resolved into two components with linear angular size 
of 24$^{\prime\prime}$ and 21-cm absorption is detected towards 
both the northern and southern components which are 
at impact parameters of 4.1\,kpc and 7.4\,kpc respectively. 
We denote such pairs as RGPs and these are not considered here. 
In the following we discuss the properties of sight lines searched 
for 21-cm absorption as part of the observations of QGPs and investigate 
their relationship with the high-$z$ 21-cm absorbers detected 
in the \mgii absorbers and DLAs towards quasars.

\subsection{Impact parameter}

\begin{figure}
\centerline{{
\psfig{figure=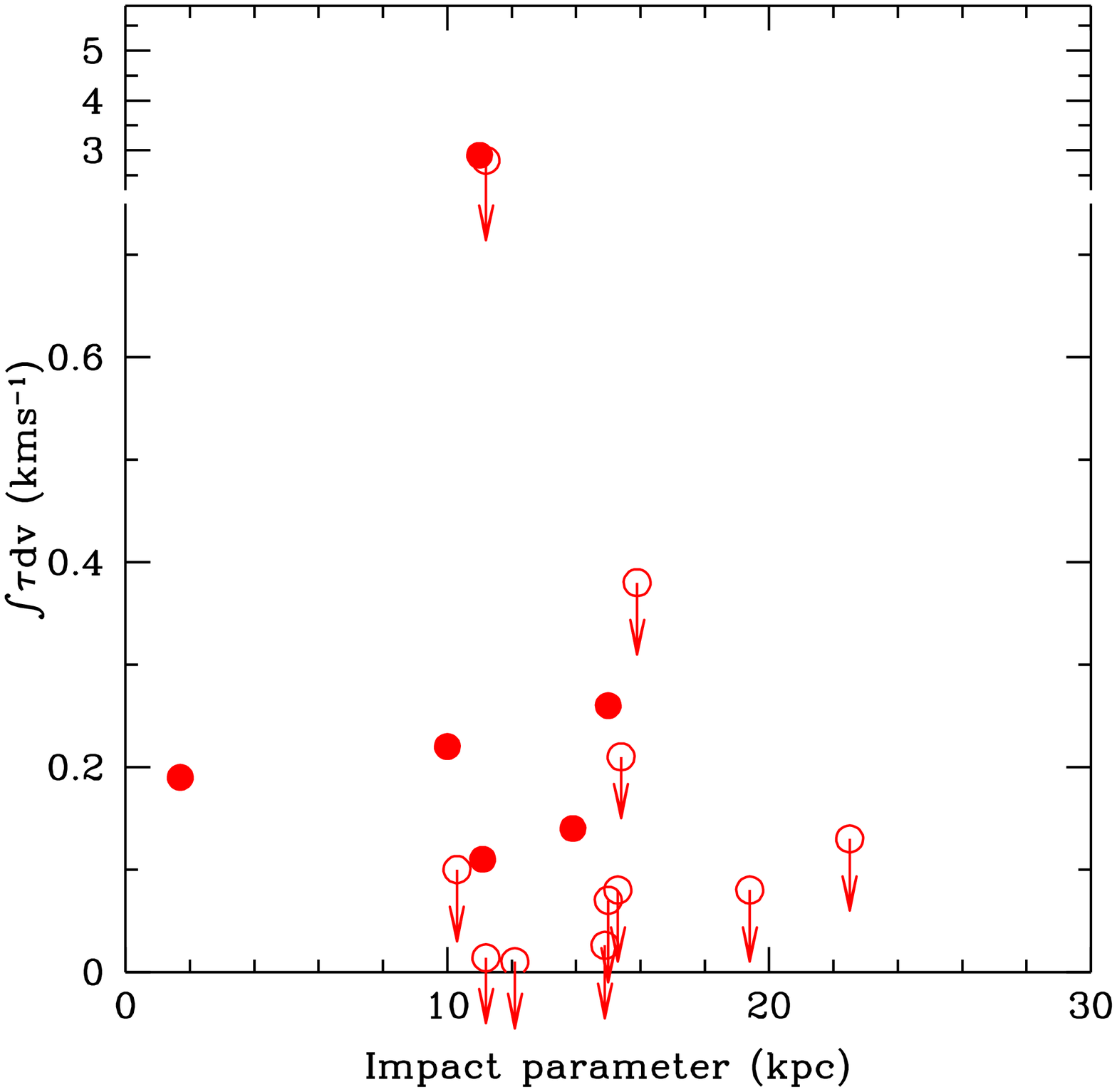,height=8.0cm,width=9.0cm,angle=0}
}}
\caption[]{Integrated 21-cm optical depth is plotted against projected 
separation between the background radio source and the centre of galaxy 
(Tables~\ref{gmrtres} and \ref{qgplit}). 
In case of non-detections values are 3$\sigma$ upper limits corresponding to 
the equivalent velocity resolution of 10\,km\,s$^{-1}$. 
Filled symbols are for the detections and open symbols are for the non-detections.  
}
\label{impact}
\end{figure}

In Fig.~\ref{impact}, we plot 21-cm integrated optical depth 
as a function of impact parameter for all the sight lines from the 
samples of QGPs (Tables~\ref{gmrtres} and \ref{qgplit}) with impact parameters 
less than 30\,kpc. We do not find any correlation between
the optical depth and impact parameter.
%

We estimate the filling factor of 21-cm absorbing 
gas for a limiting integrated  21-cm optical depth  
$\cal{T}$$_{21} (=\int\tau\,dv)$. 
When we consider QGPs with impact parameter less than 15 kpc,
we have 5 detections out of 10 sight lines where  
$\cal{T}$$_{21}\ge$\,0.1\,km\,s$^{-1}$. 
Thus the detectability of 21-cm absorption with 
$\cal{T}$$_{21}\ge$\,0.1\,km\,s$^{-1}$ is $\sim 50\%$.
This limiting value of ${\cal{T}}_{21}$ corresponds to
log\,$N$(H~{\sc i}) = 19.3, if the spin temperature of the gas is
100\,K as expected in the case of cold neutral medium.
However, the measured spin temperatures in the case of
QGPs are usually in the range 200\,K to 900\,K
(see Table~3 of Carilli \& van Gorkom 1992). So if we
take a typical value of 500\,K for the T$_{\rm S}$ then the limiting 
column density of H~{\sc i} is log\,$N$(H{~\sc i})=20.0. 
Thus  $\cal{T}\ge$\,0.1\,km\,s$^{-1}$ limit
will be sensitive to detecting similar gas clouds in high redshift
DLAs. 

\begin{table*}
\caption{Quasar-galaxy pairs with 21-cm absorption measurements from the literature 
arranged in the order of increasing galaxy redshift.}
\begin{center}
\begin{tabular}{lccccccc}
\hline
\hline
Quasar/galaxy pair    &  $z_q$  &  $z_g$            & Angular   &Distance&$\int\tau dv$ &Ref. & W$_r$(\caiia)  \\ 
                      &         &                       &Separation&        &             &      &     \\ 
                      &         &                   &($^{\prime\prime}$) & (kpc)       &      &      &     \\ 
     (1)              &   (2)   &  (3)              &   (4)     & (5)       &  (6)     &   (7)       & (8)   \\ 
\hline                                                                                           
\\
Detections \\
3C232/NGC3067             & 0.530   & 0.0049            &110.6     &11.1    &   0.11      &  1   &0.43$\pm$0.05$^+$ \\
Q1327$-$206/1327-2041     & 1.169   & 0.0180            &38.5      &13.9    &   0.14$^\dag$  &  1   &0.58$\pm$0.04$^*$ \\
Q2020$-$370/Klemola\,31A  & 1.048   & 0.0270            &18.7      &10.0    &   0.22      &  1   &0.35$\pm$0.08$^*$ \\
J104257.58+074850.5/      & 2.665   & 0.0332            &2.5       &1.7     &   0.19      & 2    &$<$1.07$^{**}$ \\ 
J104257.74+074751.3       &         &                   &          &        &                     &  & \\ 
Q0248+430/Anon            & 1.313   & 0.0520            &15.0      &15.0    &   0.26      &  3   &1.52$\pm$0.17$^*$ \\
\\
Non-detections \\
1749+70.1/NGC6503         & 0.770   & 0.0002            &321.6     &10.3    &$<$0.10$^\ddag$  &  4   & $-$              \\ 
3C309.1/NGC5832           & 0.905   & 0.00149           &372.0     &11.2    &$<$0.014     &  5   & $-$              \\
H0131+154/NGC628          & 1.330   & 0.0022            &2188.9    &94.1    &$<$0.148     &  6   & $-$              \\ 
3C275.1/NGC4651           & 0.555   & 0.0027            &220.3     &12.1    &$<$0.010     &  6   &$<$0.45$^{**}$       \\ 
0139+132/NGC660           & 0.267   & 0.0028            &911.1     &51.9    &$<$0.067     &  6   &$<$0.45$^{**}$       \\ 
2304+32B/NGC7490          & 1.937   & 0.0207            &495.8     &204.8   &$<$0.10      &  6   & $-$              \\ 
3C455/NGC7413             & 0.543   & 0.0325            &23.4      &14.9    &$<$0.026     &  6   & $-$              \\ 
2231+0953/UGC12081        & 1.854   & 0.0388            &121.6     &92.3    &$<$0.10      &  6   & $-$              \\ 
Q0446$-$208/Anon          & 1.894   & 0.0661            &12.3$^C$  &15.4    &$<$0.21      &  1   &0.57$\pm$0.06$^*$ \\ 
                          &         &                   &29.6$^E$  &37.1    &$<$0.32      &      & $-$              \\ 
                          &         &                   &28.1$^W$  &35.2    &$<$0.63      &      & $-$              \\
\\
\hline
\end{tabular}
\end{center}
\begin{flushleft}
Column 1: Quasar/galaxy pair. Column 2: Quasar redshift. 
Column 3: Galaxy redshift. 
Column 4 and 5: Angular separation and the projected separation of quasar/radio 
component from the centre of galaxy. The superscripts $C, E$ and $W$  
correspond to core, eastern and western components respectively.
Column 6: Integrated 21-cm optical depth or 3$\sigma$ upper limit in case of 
non-detections with data smoothed to 10\,km\,s$^{-1}$. 
Column 7: Reference for the 21-cm optical depth value or upper limit. 
Column 8: Rest equivalent width of \caii~ or 3$\sigma$ upper limit to it.\\ 
$^\dag$ 21-cm absorption is detected in two well-detached components separated by $\sim$250\,km\,s$^{-1}$. 
The integrated 21-cm optical depth presented here corresponds to the stronger component (see text for details).  \\
$^\ddag$ $\Delta v$=20.6\,km\,s$^{-1}$.\\ 
$^+$ W$_{\rm r}$(Ca~{\sc ii}) taken from Bowen et al (1991). \\
$^*$ W$_{\rm r}$(Ca~{\sc ii}) taken from Womble (1993). \\
$^{**}$ W$_{\rm r}$(Ca~{\sc ii}) estimated from the SDSS spectrum.\\ 
References for the H~{\sc i} absorption data$-$ 
1: Carilli \& van Gorkom (1992); 
2: Borthakur et al. (2010); 
3: Hwang \& Chiou (2004); 
4: Boiss\'e et al. (1988);
5: Haschick \& Burke (1975) and  
6: Corbelli \& Schneider (1990). 
\\
\end{flushleft}
\label{qgplit}
\end{table*}

%
Using 21-cm emission line maps of nearby galaxies, Zwaan et al. (2005)
have concluded that sight lines with log\,$N$(H~{\sc i})$\ge$ 20.3 occur
with a median impact parameter of 7.8 kpc. It is also clear from 
their Fig.~14 that when the impact parameter is between 10 and 15\,kpc, 
the probability of having high $N$(H~{\sc i}) is roughly between
50\% to 70\%. This is very much consistent with what we find here. 

\subsection{Extent of the H~{\sc i} disk of galaxy}
%
In the previous sub-section we do not consider the
possibility that the 21-cm absorption non-detection may 
simply be due to the small H~{\sc i} extent of the galaxy. 
As discussed before, the expected extent of H~{\sc i} gas  
in all but one of the galaxies in our sample (Table~\ref{gmrtres}) with no 
21-cm detection is smaller than the impact parameter. 
The only case (J124157.54+633241.6) where we have 21-cm detection 
is when the radio sight line is passing through the stellar disk 
(as we detect emission lines from the galaxy in the QSO spectrum).

Amongst the QGPs with 21-cm detections from the literature (Table~\ref{qgplit}), 
in the three cases i.e. 3C\,232/NGC3067, Q\,1327$-$206/1327$-$2041 and
Q\,2020$-$370/Klemola\,31A, the quasar sight line 
passes through the H~{\sc i} emission disk of the foreground galaxy 
$-$ a late type spiral $-$ extending well beyond the stellar disk and 
also shows signs of tidal disturbances (Carilli \& van Gorkom 1992). In the case of 
Q0248+430/G0248+430, the foreground galaxy is actually a pair 
of interacting luminous infrared galaxies.  The QSO sight line 
in the optical image passes through the tidal tail extending 
in the direction of the background quasar (Womble et al. 1990;
Hwang \& Chiou 2004). No H~{\sc i} emission was detected from
the foreground galaxy with a limiting hydrogen mass of 6$\times 10^8$ 
M$_\odot$. In the case of Quasar (J104257.58+074850.5) $-$ galaxy 
(J104257.74+074751.3) pair,    
the QSO sight line passes through the optical disk of an
H~{\sc i} deficient galaxy (Borthakur et al. 2010).
Thus all the 21-cm absorption detections are associated with the 
quasar sight lines passing through the stellar and/or extended 
H{\sc i} disk of the galaxy.

Let us now consider the 5 QGPs in Table~\ref{qgplit} with impact parameters 
less than 20\,kpc that do not show 21-cm absorption. 
In the case of 1749+70.1/NGC6503 high signal-to-noise 
H~{\sc i} emission maps show extended H~{\sc i} emission along
the semi-major axis (Greisen et al. 2009). The QSO sight line
is at an impact parameter of 10.3\,kpc along the minor axis
and not passing through the H~{\sc i} disk. In the case
of 3C275.1/NGC4651 the QSO sight line passes through the outer 
regions of the H~{\sc i} 21-cm emission disk (Schneider et al. 1993).
There is a tentative detection of 21-cm absorption noted by
Schneider et al. (1993). In the absence of further confirmation
we have considered this case as a non-detection. In the case of
3C309.1/NGC5832 and 3C455/NGC7413 no radio interferometric 
21-cm emission maps are available, whereas for the 
Q0446$-$208/Anon H~{\sc i} 21-cm emission is not detected.
It is interesting to note that in case of 
Q0446$-$208/Anon the quasar sight line passes through 
the stellar disk of the galaxy (Carilli \& van Gorkom 1992).

Thus it appears that the lack of 21-cm absorption in 50\% of
the cases with impact parameter less than 20\,kpc could be related
to the smaller extent of H~{\sc i} gas in galaxies 
and probably not related to the filling factor of 
cold gas in the extended H~{\sc i} disk/halo. Confirming
this conclusion with a large number of QGPs with low
impact parameter is important for our understanding of high-$z$
21-cm absorbers.

\subsection {Ca~{\sc ii} absorption}
\begin{figure}
\centerline{{
\psfig{figure=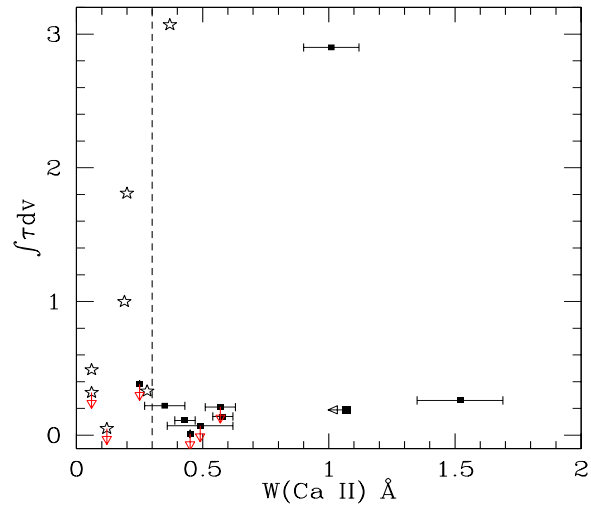,height=7.0cm,width=8.0cm,angle=0}
}}
\caption[]{The integrated 21-cm optical depth is plotted against W$_r$(\caiia) 
for QGPs (squares) and for $z\le 1$ DLAs (stars).  
}
\label{caii}
\end{figure}

The presence of Ca~{\sc ii} and/or Na~{\sc i} absorption can also be used as a 
tracer of neutral gas. Detecting significant 
amounts of these two species will ensure that the region probed is
cold and relatively well shielded from the ionizing UV radiation.
Wild et al. (2006) have detected several strong Ca~{\sc ii} absorbers 
(W$_{\rm r}$(\caiia)$\ge$0.5\AA) in the SDSS database. 
These absorbers are found to be {\sl dustier} than the systems 
detected through the presence of Mg~{\sc ii} absorption or 
DLA. 
Wild et al. (2007b) have detected strong nebular emission originating 
from the Ca~{\sc ii} absorbers by stacking methods. All these suggest that the 
QSO sight lines with Ca~{\sc ii} absorbers may be going very 
close to the star-forming regions. Indeed Zych et al. (2007)
have detected galaxies associated with four strong 
Ca~{\sc ii} absorbers with typical impact parameters less
than 23 kpc.
It is interesting to note that all the QGPs, except  
the J104257.58+074850.5/J104257.74+074751.3\footnote{It 
should be noted that the upper limit of W$_{\rm r}$(\caiia)$>$1.07\AA\ 
in this case is significantly higher than the values of W$_{\rm r}$(\caiia) 
in the case of other QGPs and DLAs with 21-cm detections 
(see Tables~\ref{gmrtres} and \ref{qgplit}), and Fig.~\ref{caii}.},  
that show 21-cm absorption also show Ca~{\sc ii} and Na~{\sc i}
absorption in the optical spectra.  Bowen et al (1991) have found
that Ca K line with a rest equivalent width $\ge$ 0.3 {\AA} is detected
only along sight lines with impact parameters less than 11 kpc.
For the cosmological parameters used here this corresponds to 
$\sim15$\,kpc.  

%
We have clear detections of Ca~{\sc ii} in two out of 5 QSO
sight lines in our sample.  In the case of two QSOs
(J084957.97+510829.0 and J111023.85+032136.1) the S/N in the
SDSS spectra does not allow us to place stringent constraint
on the rest equivalent width of Ca~{\sc ii} absorption.
In the case of J082153.82+503120.4 the limiting rest equivalent width
is good enough to rule out strong Ca~{\sc ii} absorption.
In Fig.~\ref{caii} we plot the rest equivalent width of
Ca K line against the integrated 21-cm optical depth 
for all the quasar sight lines (Tables~\ref{gmrtres} and \ref{qgplit}).
In the QGPs whenever there is 21-cm absorption detection
it is also accompanied by the detection of Ca~{\sc ii}
absorption with rest equivalent width greater than 0.3\,{\AA}.
However, the presence of strong Ca~{\sc ii} absorption alone
does not guarantee the detection of 21-cm absorption
as we found in the case of sight lines towards 
J122847.42+370612.0 (Table~\ref{gmrtres}) and Q0446-208 (Table~\ref{qgplit}).

\begin{figure}
\centerline{{
\psfig{figure=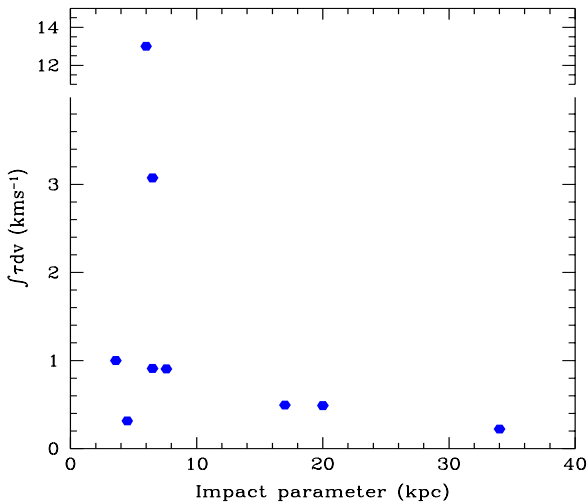,height=7.0cm,width=8.0cm,angle=0}
}}
\caption[]{The integrated 21-cm optical depth is plotted against impact parameter for $z_{abs}\lapp1$ DLAs 
with 21-cm detections from Table~4 of Rao et al. (2003).  
}
\label{lowzdla}
\end{figure}

%
The rest equivalent widths of Ca~{\sc ii} are available for 
7 DLAs at $z\le1.0$ that are searched for 21-cm absorption
(Rao et al. 2003; Kanekar \& Chengalur 2003; Nestor et al. 2008; Zych et al. 2009). 
Interestingly only one of them has Ca~{\sc ii}
equivalent width greater than 0.3 {\AA}. This is very much
different from the 21-cm detection in the case of QGPs.
There are a few interesting differences to note. First of
all, the typical impact parameters of the DLAs used in 
Fig.~\ref{lowzdla} are less than 10\,kpc 
while for the QGPs values are typically 
between 10\,kpc to 20\,kpc. The 21-cm optical depth, measured in
QGPs is typically less than what one measures
in the case of DLAs.
The low Ca~{\sc ii} equivalent width measured in the case of DLAs, 
despite having higher 21-cm optical depth and smaller impact parameters  
may be related to the low metallicity of  the high-z gas. 
Indeed Nestor et al. (2008) find that only 25\%
of the DLAs known at 0.6$\le$$z_{abs}\le$1.3 have W(Ca~{\sc ii})$\ge$0.35 \AA. 
Thus DLAs may be a biased population that avoids 
sight lines through dusty star-forming galaxies.

%
\subsection{Implications for high-$z$ 21-cm surveys}
Detectability of 21-cm absorption depends on $N$(H~{\sc i}),
${\rm T_S}$ and the fraction of the background source covered by
the absorbing gas. 
As described before, most of the 21-cm 
absorption surveys have been designed to detect 21-cm absorption
from gas pre-selected by the presence of a DLA or strong Mg~{\sc ii} absorption
(e.g. Briggs \& Wolfe, 1983; Carilli et al. 1996; Lane 2000; 
Kanekar \& Chengalur 2003, 2009a; Curran et al. 2007; Gupta et al. 2007, 2009; 
Srianand et al. 2010). In the case of DLAs, there is a possible
decline in the 21-cm detection rate with increasing redshift (Kanekar \& Chengalur 2003).
Since for the DLAs the total $N$(H~{\sc i}) along the
line of sight is known from the damped Lyman-$\alpha$ line,  
the decline in 21-cm absorption detection rate 
could be related either to the redshift evolution of spin temperature of the 
gas or to the redshift evolution of the covering factor  
(Kanekar \& Chengalur 2003; Curran \& Webb 2006; Kanekar et al. 2009b).
However these results are based on a small number of systems and 
in addition at $z\lapp$2 the DLA samples to search for 
21-cm absorption are poorly defined.

Most of the successful searches for 21-cm absorption are based
on Mg~{\sc ii} pre-selection.
A systematic survey of 21-cm absorption at $1.10\le z\le 1.45$ using 
$\sim$400~hrs of GMRT time has resulted in the largest number 
of 21-cm detections till now (Gupta et al. 2007, 2009). 
This, together with a shallower survey at 
$z\lapp1$ by Lane (2000), suggests a  possible decline in the 21-cm
absorber number density ($n_{21}$) with increasing $z$.
This is in contrast to a strong increase in the Mg~{\sc ii}
number density over the same redshift range (see
Fig.~11 of Gupta et al. 2009). 

Results presented here suggest that, while the detection 
rate of 21-cm absorption from QGPs is 50\%
when the impact parameter is less than $\sim$20\,kpc, nearly in 
every case of non-detection the quasar sight line 
does not pass through either the stellar disk or 
the extended H~{\sc i} gas. Semi-analytical model of
galaxy evolution and  direct observations from GOOD and GEMS fields
suggest a decrease in the optical size of the galaxies 
from $z\sim 0$ to $z\sim 2$ (Mo et al. 1998,
Ravindranath et al. 2004; Trujillo et al. 2006). The 
observations are consistent with the size of
a galaxy at a given stellar mass at $z\sim 1.3$ 
to be $\sim 70$\% that of a galaxy at $z\sim 0$. 
Semi-analytic models coupled with Millennium dark matter 
simulations suggests that the radius of the H~{\sc i} gas
at $z\sim 1$ is roughly half of that at $z\sim 0$ (Obreschkow 
\& Rawlings, 2009). Based on this we expect the impact 
parameter around a typical galaxy at $z\sim1.3$ to be 
a factor 1.5 to 2 less than the value of $\sim$20\,kpc we find for the 
QGPs and $z\lapp1$ DLAs (Rao et al. 2003).
Therefore, the observed decrease in $n_{21}$ could be a 
consequence of the small size of the optical disk and the 
extent of associated cold H~{\sc i} disk/halo. 
This needs to be understood in the context of the observed 
increase in the number density of absorption systems with 
redshift at optical/UV wavelengths.
Obviously, this requires identifying both (1) the physical processes 
that determine the absorption cross-section and (2) the 
underlying population of absorbing galaxies.  
The SDSS database provides the means to make progress in this 
direction for 21-cm absorbers by constructing systematic samples 
of QGPs and RGPs.

Spectral stacking methods have shown that high equivalent width
Mg~{\sc ii} systems are possibly associated with  star forming
regions (Noterdaeme, Srianand \& Mohan, 2010; M\'enard et al. 2009).
It is also found that the average emission line flux for a given 
equivalent width range depends only weakly on the redshift 
(M\'enard et al. 2009). All this suggests that 
stacked spectra of Mg~{\sc ii} systems with 21-cm 
absorption should show stronger emission lines than the ones
without 21-cm absorption. It will be interesting to check this
once we have a sufficient number of 21-cm absorbers.

\section{summary}

We have presented the results of our GMRT {\sl mini}-survey to search for 
21-cm absorption in a sample of 5 QGPs (total 9 sight lines) at $0.03\le$$z\le0.18$.  
We report one clear detection of 21-cm absorption in the quasar 
($z_q$=2.626 SDSS J124157.54+633241.6) $-$ ($z_g$=0.1430 SDSS J124157.26+633237.6). 
In this case the quasar sight line pierces through the stellar disk of the galaxy.
We infer metallicity, star-formation rate and reddening for the galaxy
from the SDSS spectrum. The column density inferred from the reddening
estimates are consistent with the 21-cm absorbing gas being cold.

Combining our sample with the $z\le0.1$ QGPs (10 additional 
sight lines with impact parameter less than 20\,kpc) available
from the literature, we show the detectability of 21-cm absorption with
integrated optical depth in excess of 0.1 km/s to be $\sim 50$\%
when the impact parameter is less than $\sim$20\,kpc. 
Using the surface brightness profiles and
well established relationship between the optical size and extent of 
the H~{\sc i} disk (inferred from 21-cm emission observations)
known for {\sl nearby} galaxies, we conclude that in most 
of the cases of 21-cm absorption non-detection, 
the sight lines may not be passing through the H~{\sc i}
gas.

We also find that whenever 21-cm absorption is seen Ca~{\sc ii}
absorption is detected in the optical spectrum. However, the reverse
is not true. This means even at low redshifts metal absorption may
originate from a larger area where the H~{\sc i} gas may be warm 
(high ${\rm T_S}$) or ionized. 
Further we notice that $z<1$ DLAs with 21-cm absorption detections 
have lower \caii equivalent widths despite having higher 
21-cm optical depths and smaller impact parameters.  
This suggests that the current sample of DLAs may be a biased population that 
avoids sight lines through dusty star-forming galaxies.  

At present the 
sample size is small and H~{\sc i} 21-cm emission maps are not available
for most of the objects in the sample to investigate the dependence
of detectability on various quantities like (i) extent of the H~{\sc i} disk, 
(ii) metallicity and star-formation rate, (iii) dependence of 21-cm 
absorption on the type of galaxies, and (iv) the role of 
environmental factors such as tidal interactions and merger effects.  
The SDSS database provides the means to perform a systematic study to 
investigate these issues and understand the nature of 21-cm absorbers.

\section{Acknowledgements}
We thank the referee, Vivienne Wild, for the very useful and 
detailed comments. 
We wish to thank Vijay Mohan and Swara Ravindranath for useful discussions.
We thank GMRT staff for their co-operation during the observations.
GMRT is run by the National Centre for Radio Astrophysics of 
the Tata Institute of Fundamental Research.
Galaxy spectra obtained at the Apache Point Observatory were made 
using the 3.5-meter telescope, which is owned and operated by the
Astrophysical Research Consortium.
We acknowledge the use of SDSS spectra from the archive at http://www.sdss.org/.
Funding for the SDSS and SDSS-II has been provided by the Alfred    
  P. Sloan Foundation, the Participating Institutions, the National   
  Science Foundation, the U.S. Department of Energy, the National     
  Aeronautics and Space Administration, the Japanese Monbukagakusho, the
  Max Planck Society, and the Higher Education Funding Council for     
  England.
D.V.B is funded through NASA LTSA grant NNG05GE26G.



\end{document}